\documentclass[conference]{IEEEtran}
\IEEEoverridecommandlockouts
\usepackage{cite}
\usepackage{amsmath,amssymb,amsfonts}
\usepackage{graphicx}
\usepackage{textcomp}
\usepackage{xcolor}
\usepackage{bbm,bm}
\usepackage{svg}
\usepackage[ruled]{algorithm}
\usepackage[algo2e]{algorithm2e} 
\usepackage{algorithmic}
\usepackage{ragged2e}
\usepackage[nonumberlist, style=long, toc, acronym, symbols]{glossaries}
\usepackage{multirow}
\usepackage{colortbl}
\usepackage{tabulary}
\usepackage{etoolbox}
\usepackage{subcaption}  %subfigure
\usepackage{tcolorbox}
\usepackage{empheq}
\usepackage[normalem]{ulem}

\begin{document}

% Network related glossary
\newacronym{sdwan}{SD-WAN}{Software Defined-Wide Area Network}
\newacronym{wan}{WAN}{Wide Area Network}
\newacronym{mpls}{MPLS}{Multi-Protocol Label Switching}
\newacronym{qos}{QoS}{Quality-of-Service}
\newacronym{vpn}{VPN}{Virtual Private Networks}
\newacronym{sdn}{SDN}{Software Defined Networking}
\newacronym{nfv}{NFV}{Network Function Virtualisation}
\newacronym{ar}{AR}{Access Router}
\newacronym{ip}{IP}{Internet Protocol}
\newacronym{lb}{LB}{Load Balancer}
\newacronym{tcp}{TCP}{Transmission Control Protocol}
\newacronym{mlu}{MLU}{Maximum Link Utilization}
\newacronym{lu}{LU}{Link Utilization}
\newacronym{cpe}{CPE}{Customer Premises Edge}
\newacronym{sla}{SLA}{Service Level Agreement}
\newacronym{ecmp}{ECMP}{Equal Cost Multi-Path}
\newacronym{api}{API}{Application Programming Interface}
\newacronym{nlp}{NLP}{Non Linear Programming}

% Learning related glossary
\newacronym{cpu}{CPU}{Central Processing Unit}
\newacronym{gpu}{GPU}{Graphics Processing Unit}
\newacronym{cuda}{CUDA}{Compute Unified Device Architecture}
\newacronym{ml}{ML}{Machine Learning}
\newacronym{drl}{DRL}{Deep Reinforcement Learning}
\newacronym{dl}{DL}{Deep Learning}
\newacronym{rl}{RL}{Reinforcement Learning}
\newacronym{cpo}{CPO}{Constrained Policy Optimization}
\newacronym{rcpo}{RCPO}{Reward Constrained Policy Optimization}
\newacronym{cbf}{CBF}{Control Barrier Function}
\newacronym{mdp}{MDP}{Markov Decision Process}

\newacronym{ddpg}{DDPG}{Deep Deterministic Policy Gradient}
\newacronym{td3}{TD3}{Twin-Delayed Deep Deterministic Policy Gradient}
\newacronym{trpo}{TRPO}{Trust Region Policy Optimization}
\newacronym{ppo}{PPO}{Proximal Policy Optimization}
\newacronym{scip}{SCIP}{Solving Constraint Integer Programs}
%Math
\newglossaryentry{As}{%
name=\ensuremath{\mathcal{A}^s},
description={set of safe actions}
}

\newglossaryentry{Ap}{%
name=\ensuremath{\mathcal{A}^p},
description={set of predicted actions}
}

\newglossaryentry{acbf}{%
name=\ensuremath{a^{CBF}},
description={safe action}
}

\newglossaryentry{arl}{%
name=\ensuremath{a_\theta},
description={predicted action}
}

\newglossaryentry{pk}{%
name=\ensuremath{\mathbb{P}_k},
description={set of available paths in tunnel k \in K}
}

\newglossaryentry{de}{%
name=\ensuremath{d_e},
description={delay on each edge/link e \in p \forall p\in \mathbb{P}_k }
}

\newglossaryentry{dpk}{%
name=\ensuremath{d_p^k},
description={delay on each path p of tunnel k }
}

\newglossaryentry{dk}{%
name=\ensuremath{d^k},
description={delay of tunnel k }
}

\newglossaryentry{Dtk}{%
name=\ensuremath{T^k},
description={Source traffic of tunnel k }
}

\newglossaryentry{Dtkh}{%
name=\ensuremath{\widehat{T_t^k}},
description={Admitted traffic of tunnel k }
}

\newglossaryentry{xpk}{%
name=\ensuremath{{x_p^k}},
description={split ratio on path p of tunnel k }
}

\newglossaryentry{lpk}{%
name=\ensuremath{{l_p^k}},
description={load on path p of tunnel k }
}

\newglossaryentry{bpk}{%
name=\ensuremath{b_p^k},
description={boolean variable indicating flow k has traffic routed through path p }
}

\newglossaryentry{mlue}{%
name=\ensuremath{\mu_e},
description={\acrfull{mlu} of link e}
}

\newglossaryentry{lue}{%
name=\ensuremath{\nu_e},
description={\acrfull{lu} of link e}
}

\newglossaryentry{ce}{%
name=\ensuremath{c_e},
description={capacity of link/edge e }
}

\newglossaryentry{xv}{%
name=\ensuremath{\bm{x}},
description={weight for link i}
}

\newglossaryentry{N}{%
name=\ensuremath{\mathcal{N}},
description={set of local links attached in WAN network}
}

\newglossaryentry{exp}{%
name=\ensuremath{\mathcal{D}},
description={experience array}
}

\title{Towards Safe Load Balancing based on Control Barrier Functions and Deep Reinforcement Learning
}

\author{Lam Dinh, Pham Tran Anh Quang, J\'er\'emie Leguay\\
Huawei Technologies Ltd., Paris Research Center, France  \\
\{lam.dinh, phamt.quang, jeremie.leguay\}@huawei.com\}
}

\maketitle

\begin{abstract}
\acrfull{drl} algorithms have recently made significant strides in improving network performance. Nonetheless, their practical use is still limited in the absence of safe exploration and safe decision-making. 
In the context of commercial solutions, reliable and safe-to-operate systems are of paramount importance. 
Taking this problem into account, we propose a safe learning-based load balancing algorithm for \acrfull{sdwan}, which is empowered by \acrfull{drl} combined with a \acrfull{cbf}. It safely projects unsafe actions into feasible ones during both training and testing, and it guides learning towards safe policies. We successfully implemented the solution on \acrshort{gpu} to accelerate training by approximately 110x times and achieve model updates for on-policy methods within a few seconds, making the solution practical. We show that our approach delivers near-optimal \acrfull{qos} performance in terms of end-to-end delay while respecting safety requirements related to link capacity constraints. We also demonstrated that on-policy learning based on \acrfull{ppo} performs better than off-policy learning with \acrfull{ddpg} when both are combined with a \acrshort{cbf} for safe load balancing. 
\end{abstract}


\section{Introduction}
Many enterprises are adopting \acrfull{sdwan}~\cite{yang2019software}  technologies to trade-off between cost-effectiveness and \acrfull{qos} satisfaction. Relying on a network overlay, this architecture allows businesses to interconnect multiple sites (enterprise branches, headquarter, data centers) without the need to deploy their own physical infrastructure, making it cost effective. A centralized controller maintains a set of policies deployed at access routers which send traffic
to their peers over several transport networks (e.g., private lines, broadband internet, 5G).
Typically, access routers are responsible for enforcing traffic engineering, for instance load balancing,
and queuing policies to meet \acrfull{sla} requirements in terms of
end-to-end \acrfull{qos}, security, etc. At a slow
pace, the controller maintains policies, while access devices
make real-time decisions for every flow. \acrshort{sdwan} provides a cost efficient alternative to private lines and drastically improves \acrshort{qos} compared to best effort solutions such as \acrfull{vpn}. 

Several solutions have been proposed for load balancing in \acrshort{sdwan} networks to satisfy \acrshort{sla} requirements. Centralized and distributed path selection mechanisms have been proposed to maximize an utility function~\cite{magnouche2021distributed} or \acrshort{sla} satisfaction using a performance model~\cite{quang2022intent}. To optimize latency and other \acrshort{qos} parameters for a specific set of flows, closed-form performance models, using network calculus or queuing models, can be embedded into routing optimization algorithms. For instance, authors in \cite{benAmeur2006} considered the Kleinrock function~\cite{kleinrock2007communication} to minimize latency. However, accurate analytical models for end-to-end \acrshort{qos} performance metrics can be difficult to derive and integrate into routing optimization algorithms. Indeed, unknown scheduling parameters, behaviors inside the \acrshort{wan} and transport-layer mechanisms are, in practice, too difficult to capture by tractable performance models. The challenge to integrate accurate analytical
models makes \acrshort{qos} routing and lod balancing a great opportunity for
model-free solutions.

Instead of having a predefined performance model, \acrfull{rl} agents
can interact with the environment and evaluate the outcomes of their actions thanks to a reward function.
\acrfull{drl}~\cite{xu2018experience} combines \acrfull{dl} and \acrfull{rl} principles (e.g., parameterize policies with neural networks). It has been first applied for routing to optimize network utility
under the umbrella of \textit{experience-driven networking}~\cite{xu2018experience}. Since, several single-agent and multi-agent \acrshort{drl} solutions have been proposed to tune queues and load balancing policies to satisfy \acrshort{qos} requirements or minimize congestion~~\cite{mai2020multi, kim2021deep, houidi2022constrained, fawaz2023graph, troiaDeepReinforcementLearning2021, RILNET}. However, even if \acrshort{drl} has demonstrated a tremendous potential for improving \acrshort{sdwan} performance,
most of the literature only focuses on off-policies and their performance once training has converged, without paying attention to safety during both learning and testing. In addition, as  network environments often drift, frequent retraining of the full model to adapt to changes can be cumbersome. In \acrshort{drl}, the trial-and-error process is crucial for environment exploration and learning but, at the same time, unsafe actions cannot be deployed as they degrade network quality in production systems. For example, a poor load balancing policy might cause severe congestion and accidentally degrades the overall network performance.  
Therefore, considering safety during both training phase, in particular for more practical on-policies, and testing is key for the wide adoption of \acrshort{rl} algorithms in network systems.

Taking those issues into consideration, this work seeks to complement current \acrshort{drl}-based load balancing solutions with an additional safety shield.
Based on the safe learning approach primarily presented in \cite{chengEndtoEndSafeReinforcement2019}, which is designed for critical robotic systems, we propose a safe load balancing solution for \acrshort{sdwan}. Our solution achieves 
near-optimal \acrfull{qos} performance in terms of average end-to-end delay, while safety concerns related to the violation of link capacity constraints are fully considered. 

The contributions of our work are the following. First, we describe the target \acrshort{sdwan} system and formulate a load balancing problem to minimize the average tunnel latency that can efficiently be solved with \acrshort{rl}. Then, we design a dedicated \acrfull{cbf} based on local search to deliver safety on top of gradient-based \acrfull{drl} algorithms (e.g., off/on policy learning). 
Finally, we evaluate our \acrshort{drl}-\acrshort{cbf} solution in both training and testing phases. We compare our solution to traditional learning algorithms (e.g., \acrshort{ddpg}, \acrshort{ppo}) where safety is only handled in the reward function, without any strict guarantees. We show that our solution can minimize latency while providing full safety guarantees. In a controlled environment, we also demonstrate that the \acrshort{qos} obtained is very close to the optimal solution derived from a non linear integer model solved with the \acrshort{scip} \cite{bestuzhevaSCIPOptimizationSuite2021} solver. In addition, in terms of execution time, we implemented \acrshort{drl}-\acrshort{cbf} algorithms on  \acrshort{gpu} and managed to accelerate training by approximately 110x times and achieve model updates for on-policy methods within a few seconds, making the full solution practical.

The paper is organized as follows. Related work is discussed in Section~\ref{sec:related}. Section~\ref{sec:system} presents the system model and formulates the load balancing problem. Section~\ref{sec:CBF} describes the proposed solution, whereas Section~\ref{sec:results} provides numerical results, demonstrating our proposed algorithm performance. Finally, Section~\ref{sec:conclusion} concludes the paper.

\section{Related work}
\label{sec:related}

\acrfull{drl} algorithms for \acrshort{sdwan} controllers has been proven to improve overall network performance~\cite{mai2020multi, kim2021deep, houidi2022constrained, fawaz2023graph, troiaDeepReinforcementLearning2021, RILNET}. For instance, Troia et al. \cite{troiaDeepReinforcementLearning2021} have shown a target \acrshort{qos} can be achieved through the proper design of the reward function. Similar results have been achieved using multiple agents~\cite{houidi2022constrained}.
However, as mentioned before, most of the literature only focuses on 1) off-policies and 2) network performance  without paying attention to safety.

Indeed, RL-based load balancing systems may violate capacity constraints both in training and testing phases. When globally minimizing the end-to-end delay, unsafe actions creating congestion and violating  capacity may be taken to get rid of portions of the traffic and improve the reward. To deal with such abnormal behaviors, the LearnQueue~\cite{bouacidaPracticalDynamicBuffer2019} reward has been introduced to minimize the end-to-end delay while penalizing traffic rejections.
However, it requires to weight properly the two objectives, which can be tedious in practice since it heavily depends on the environment.
To address these limitations and avoid manual parameter tuning, the first work to systematically optimize \acrshort{qos} under safety constraints for load balancing has been presented by Kamri et al.~\cite{kamriConstrainedPolicyOptimization2021b}.
This work applies the \acrfull{rcpo} algorithm\cite{tesslerRewardConstrainedPolicy2018} where the reward integrates traffic rejection as a constraint using  Lagrangian relaxation. During training, the algorithm finds the optimal Lagrangian multiplier. Following this work, Zhang et al. 
\cite{zhangPathPlanningModel2022} investigate a path planning problem based on constrained policy iteration to improve the performance of multiple path selection. A safe load balancing strategy for ultra-dense network is also discussed by Huang et al. \cite{huangProactiveLoadBalancing2022}. In this work, they proposed a proactive load balancing algorithm on top of \acrfull{cpo} \cite{achiamConstrainedPolicyOptimization2017a}, which has been proven to guarantee optimal policy under (safety) constraints. 

Although several papers have presented \acrshort{drl} algorithms for load balancing ~\cite{troiaDeepReinforcementLearning2021,  bouacidaPracticalDynamicBuffer2019} with some constraints\cite{kamriConstrainedPolicyOptimization2021b,huangProactiveLoadBalancing2022}), they 1) all provide \textit{soft} guarantees during exploration  and 2) mostly ensure safety during exploitation. To mitigate these issues, we propose to employ a \acrfull{cbf}~\cite{amesControlBarrierFunctions2019}  on top of current \acrshort{drl} algorithms for (safe) load balancing optimization with \textit{hard} guarantees.

\section{System model}
\label{sec:system}

\begin{figure}[ht]
    \centering
    \includegraphics[scale=0.25]{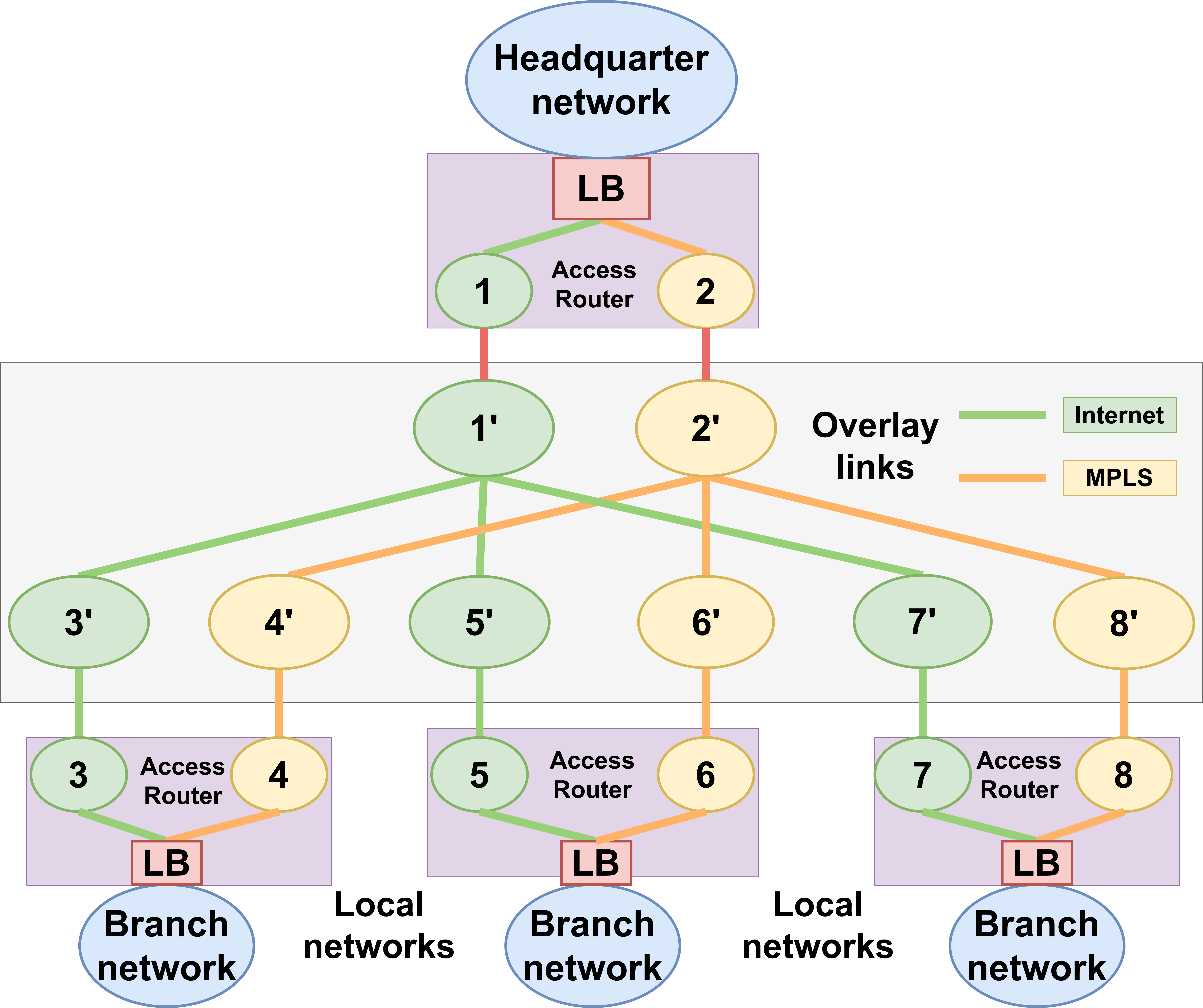}
    \caption{SD-WAN network with an headquarter and 3 branches.}
    \label{fig:scenario}
\end{figure}

Figure~\ref{fig:scenario} presents a typical \acrshort{sdwan} use case where the headquarter and 3 branches of an enterprise are interconnected either via an Internet connection and a \acrfull{mpls} private line. Traffic is issued by applications at both headquarter and branches. $6$ OD (Origin-Destination) flows, also called \textit{tunnels}, are considered, one per headquarter and branch pair in each direction. Each tunnel has two paths for Internet and \acrshort{mpls}. A \acrfull{lb} agent at each \acrfull{ar} splits the traffic according to the policy received by the centralized controller.

This overlay network can be modeled as a graph where core links represent overlay links and edge links correspond to \acrshort{wan} ports at \acrshort{ar} routers. As depicted in Figure \ref{fig:scenario}, links $N'-N$ at ports $N$ are used to model the egress / ingress capacity provisioned at each transport network. In practice, \acrshort{wan} ports can receive traffic from multiple sites, potentially of higher capacity, and they might be congested in high load conditions. Let formally consider a graph $G=(V,E)$ where $V$ is the set of nodes and $E$ is the set of edges.
Each tunnel $k$ in a set of tunnels $K$ can use a set of \textit{candidate paths} denoted as  $\gls{pk}$ (e.g., Internet and \acrshort{mpls}) to load balance traffic. Each edge $e$ carries an instantaneous load $l_e$ and has a capacity $\gls{ce}$. Let denote  $\gls{Dtk}$ % and \jeremie{used somwhere after?:} $\gls{Dtkh}$ 
 the traffic demand %and the amount of admitted traffic 
of tunnel \textit{k}  at time \textit{t}. Each \acrshort{lb} agent applies at time $t$ a split ratio $\gls{xpk}$ for each tunnel $k$ over each path $p \in \gls{pk}$ 
($\gls{xpk} \in [0,1]$ and $\sum_{p \in \gls{pk}} \gls{xpk}=1\ \forall k \in K$).

The delay on each path $p$ for a tunnel $k$ is denoted $\gls{dpk}$ and the tunnel delay, denoted $\gls{dk}$ is computed as follows:
\begin{equation}
    \gls{dk} = \underset{p \in \gls{pk} }{\max} \quad \gls{dpk}
\end{equation}

In practice, $\gls{dpk}$ is continuously measured by access routers. \\

\textbf{Problem formulation.} The main objective is to derive an optimal load balancing policy so that the \acrshort{sdwan} overlay delivers the best \acrshort{qos}. In the rest of the paper, we consider as primary target the  minimization of the average tunnel delay under the constraint that link 
capacity constraints are not violated (safety constraint). Indeed, this safety measure prevents that 1) congestion is induced by miss configured split ratios and that 2) the average tunnel delay is not artificially minimized by rejecting traffic (i.e., by intentionally creating some congestion). 
To prevent high link delays, or very heterogeneous ones even if the average delay is low, a common practice is to enforce a \acrfull{mlu} $\mu \in [0,1]$ over all links.
\acrshort{lb} agents at headquarter and branches are configured by a network controller. 

In this case, load balancing policies, which are derived according to dynamic tunnel traffic \gls{Dtk}, solve the following optimization problem: 

\begin{align*}
    \underset{\gls{xpk}}{\min} ~~~~~~~&   \frac{\sum_{k\in K}\gls{dk} }{K}\tag{$\mathcal{P}$} \label{eq:P0}\\
     \text{s.t.}~~~~~& \sum_{k \in K}  \sum _{i=0, p \in \gls{pk}} ^{|p|} \gls{Dtk}.x^k_i \leq \mu . c_e ~~~~~ \forall e \in E \tag{$\mathcal{C}_0$} \label{eq:C0} \\
     & \sum _{i=0} ^{|p|} x^k_i =1  ~~~~~~~~~~~~~~~~~~~~~~~~\forall x^k_i \in [0,1]\tag{$\mathcal{C}_1$} \label{eq:C1} \\
%     & 0 \leq \gls{mlue} \leq 1 & \forall e \in E \tag{$\mathcal{C}_2$} \label{eq:C2} \\
\end{align*}

where problem 
\eqref{eq:P0} minimizes the average tunnel delay. Constraints \eqref{eq:C0} guarantee that the traffic over each edge $e$ in the network is kept under the \acrshort{mlu}. 
Constraints \eqref{eq:C1} ensure that splits ratios sum to $1$.

\section{Learning-based and safe load balancing}
\label{sec:CBF}

In this section, we propose a constrained policy optimization for load balancing based on  \acrfull{drl} algorithms and a \acrfull{cbf} \cite{chengEndtoEndSafeReinforcement2019}. 

\subsection{Learning-based optimization}

Our optimization problem can be formulated as a \acrfull{mdp} which is defined by the tuple  $\left<\mathcal{S},\mathcal{A},\mathcal{R},\mathcal{T},\gamma  \right>$ where $\mathcal{S}$ represents the set of states, $\mathcal{A}$ is the set of available actions, $\mathcal{R}: \mathcal{S} \times \mathcal{A} \times \mathcal{S} \rightarrow \mathbb{R}$ is the reward function which gives the reward for the transition from one state to another given an action, $\mathcal{T}: \mathcal{S} \times \mathcal{A} \times \mathcal{S} \rightarrow [0,1]$ is the transition matrix, which gives the probabilities of transitioning from one state to another given an action 
and $\gamma \in [0,1]$ is the discount factor. A policy $\pi: S \rightarrow  A$ refers to the probability of taking an action $a \in \mathcal{A}$ under state $s \in \mathcal{S}$. The agent iteratively interacts with the (networking) environment to learn an optimal policy $\pi^*$ that chooses actions with the best payoff. To solve the \acrshort{mdp} associated with problem \eqref{eq:P0}, a centralized controller is employed in the network.  The controller can periodically collect at every time $t$ information from the different \acrshort{ar} routers: the traffic demand $\gls{Dtk}$, the delay $\gls{dk}$ of each tunnel $k\in K$ and the maximum link utilization $\mu$. In this context, we consider the observation space or state $s_t$ as the set of traffic demands $\gls{Dtk}$ for all tunnels $k \in K$. The action space is determined as the set of split ratios $\gls{xpk}$ for all paths $p \in \gls{pk}$ of each tunnel $k \in K$.

By using \acrshort{rl} algorithms, we learn the optimal stochastic control policy $\pi(a|s)$ that maximizes the performance measure (i.e., $J(\pi)$) which is expressed as follows:
\begin{equation}
    J(\pi) = \mathbb{E}_{\tau \sim \pi} \left[ \sum_t ^\infty \gamma^t r_t(s_t,a_t) \right] 
\end{equation}
where $\tau \sim \pi$ denotes a trajectory during which actions are sampled according to the policy $\pi(a|s)$. 

Given the fact that policy gradient  \acrshort{rl} methods have shown good performance in continuous control problem \cite{silverDeterministicPolicyGradient2014} \cite{suttonReinforcementLearningIntroduction2018}, in the rest of this paper, we consider off-policy learning (e.g. \acrfull{ddpg}, \acrfull{td3}) and on-policy learning (e.g., \acrfull{ppo}) algorithms. In the former technique, actions are sampled to encourage the agent to explore the environment. Then, a target deterministic policy is derived from these actions, supported by the actor-critic architecture \cite{silverDeterministicPolicyGradient2014} with a replay buffer \cite{andrychowiczHindsightExperienceReplay2018}. On the other hand, on-policy methods iteratively derive the target policy based on samples obtained from their own previous actions and the update is carried out by solving the following optimization problem: 
\begin{align}
    & \pi_{i+1} = \underset{\pi}{argmax}  \sum_s \rho_{\pi_i} (s) \sum_a \pi(a|s) A^{\pi_i}(s,a) \\
    & \text{s.t.} ~~~~ KL\left[\pi_{i+1}(.|s_t),\pi_{i}(.|s_t) \right]  \leq \delta_{KL}
\end{align}
where $A^{\pi_i}(s,a)$ is the advantage of performing action $a$ (sampled by old policy $\pi_i$) at state $s$, $\rho_{\pi_i} (s)$ represents the frequency of visit of state $s$ under policy $\pi_i$  and $KL\left[\pi_{i+1}(.|s_t),\pi_{i}(.|s_t) \right]$ refers to Kullback-Leibler divergence between new policy and old policy \cite{Kullback51klDivergence} and this value is bounded by a target $\delta_{KL}$. Both on and off-policy gradient algorithms do not consider safety, therefore our goal is to complement these model-free learning algorithms with a \acrfull{cbf} which guarantees safety during exploration and exploitation. 

\begin{figure}[ht!]
\centering
\includegraphics[scale=0.22]{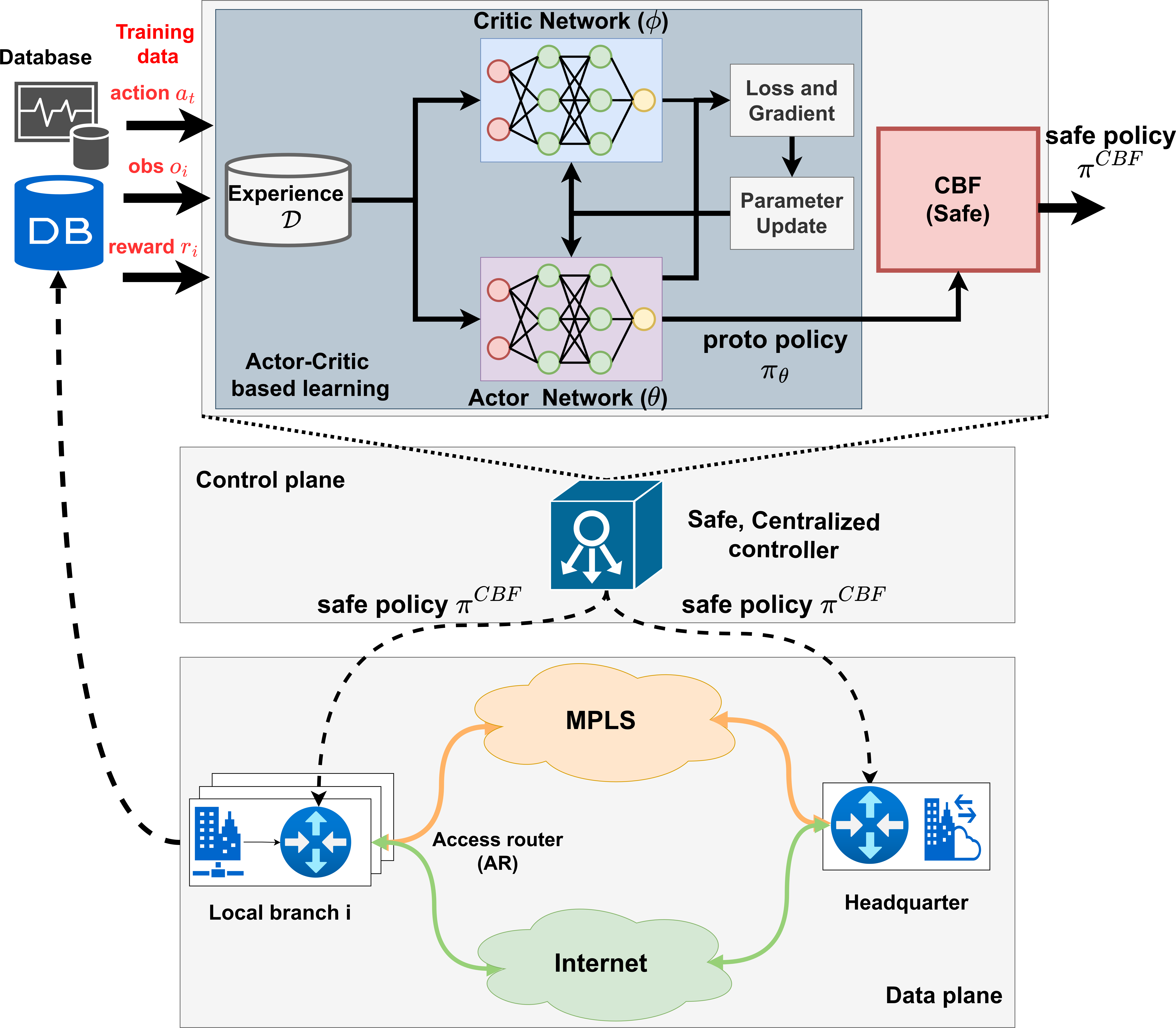}
\caption{Safety-based actor-critic learning architecture.}
\label{fig:central_dm}
\end{figure}

Figure \ref{fig:central_dm} introduces the control plane architecture where each \acrshort{lb} agent is centrally configured by the controller. The controller can periodically query the network state (i.e., tunnel delays, OD traffic, MLU) at each \acrshort{lb} and its control system is composed by two main blocks that focus on policy optimization and safety, respectively. The former block, which is based on an actor-critic  architecture\cite{silverDeterministicPolicyGradient2014}, learns the (unconstrained) optimal policy. 
Accordingly, the value function and the policy are approximated using two different neural networks, parameterized by $\phi$ and $\theta$, respectively. Each learning parameter is responsible for either maximizing the actor's objective function (i.e., $\mathcal{J}_b^A$) or the critic's loss function (i.e., $\mathcal{L}_b^C$). The specific expressions for $\mathcal{J}_b^A$ and $\mathcal{L}_b^C$ heavily depend on the off/on policy reinforcement learning algorithms being used. For instance, the explicit functions for \acrshort{ddpg} (off-policy) and \acrshort{ppo} (on-policy) algorithms  can be respectively found in \cite{silverDeterministicPolicyGradient2014} and \cite{schulmanProximalPolicyOptimization2017}.
The details of the optimization procedure are shown in Algorithm \ref{alg:rlAlgo}.

In order to enforce safe exploration during learning and safe policy execution during testing, a safety block, based on a \acrshort{cbf} function~\cite{amesControlBarrierFunctions2019}, is implemented on top of \acrshort{drl} algorithms.

\begin{algorithm}[tp]
\small
\caption{RL-based optimization algorithm}\label{alg:rlAlgo}
Initialize  parameterized \acrshort{rl} policy $\pi_{\theta_0}$ \\
Initialize value function parameter $\phi_0$ \\
Define episode length $L_{eps}$, batch size $B$, total episodes $N_{eps}$\\
Initialize experience array $\gls{exp}=\{\oslash\}$\\
\If{episode $k < N_{eps}$}{
\For {time step $t=1,...,L_{eps}$}{
Observe state $s_t$ \\
Sample action $a_t \sim \pi_{\theta_k}(.|s_t)$ \\
Perform local search algorithm \\
\begin{equation*}
{\colorbox{gray}{$\gls{acbf}_t = Local\_Search(s_t,a_t)$}}
\end{equation*}

Deploy action $\gls{acbf}_t$ and obtain reward $r_t$, next state $s_{t+1}$\\
Store experience tuple $<s_t,\gls{acbf}_t,s_{t+1},r_t>$ in $\gls{exp}$\\
}

\For {batch sampling $B$ experiences in \gls{exp}}{
Perform gradient \textbf{ascent} on actor network\\
\begin{equation*}
    \theta_{k+1} \leftarrow \frac{1}{|B|}\underset{\theta_k}{\arg \max} \sum _{b=0}^{|B|} \mathcal{J}_b^A(\theta_k)
\end{equation*}
Perform gradient \textbf{descent} on critic network\\
\begin{equation*}
    \phi_{k+1} \leftarrow \frac{1}{|B|}\underset{\phi_k}{\arg \min} \sum _{b=0}^{|B|} \mathcal{L}_b^C(\phi_k)
\end{equation*}
}
}
\textbf{return} parameterized policy $\pi_\theta$, parameterized value function $\phi$ 
\end{algorithm}

\subsection{Safe policy exploration and exploitation}

\begin{figure}[b]
    \centering
    \includegraphics[scale=0.2]{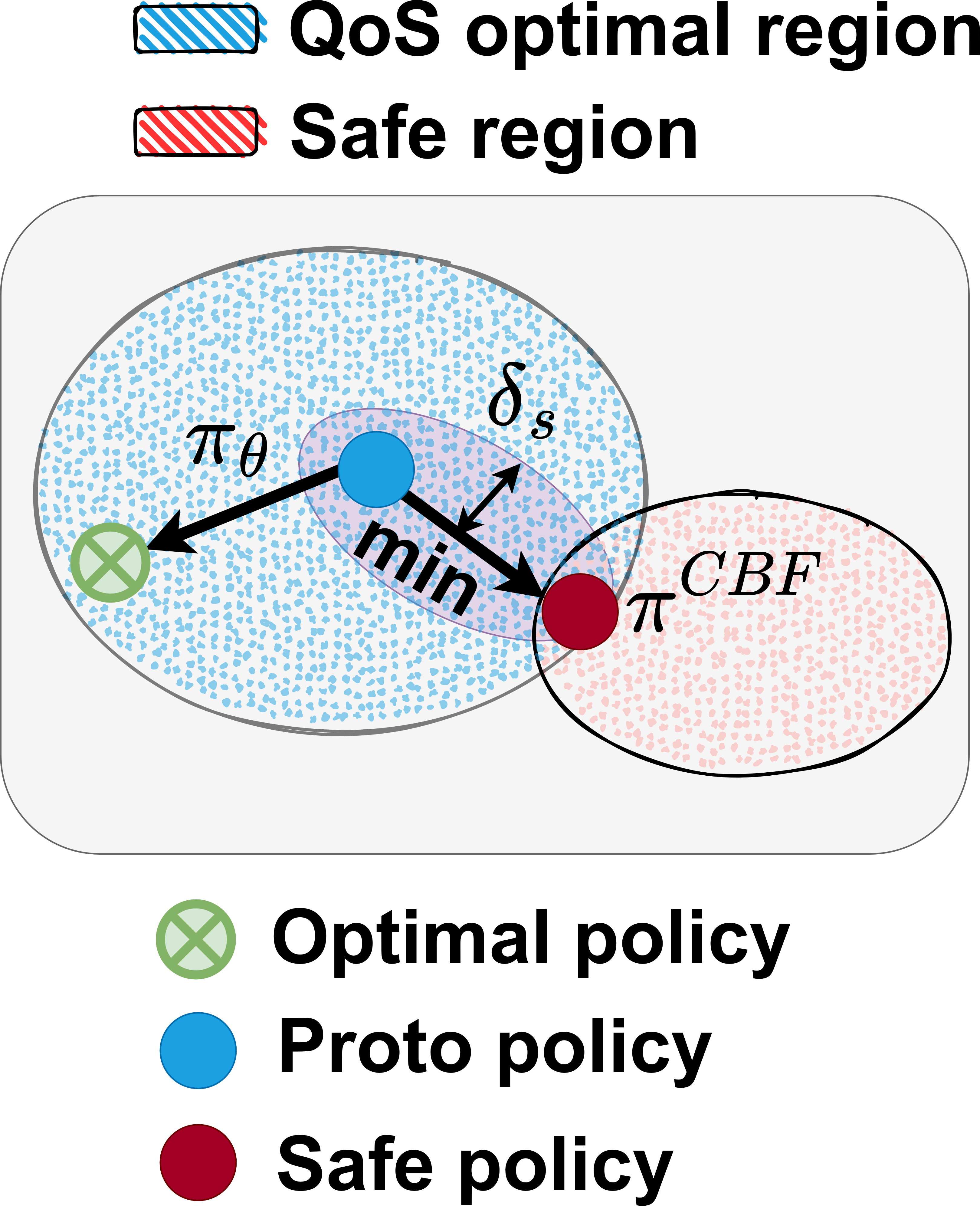}
    \caption{From proto-policy to safe policy with CBF.}
    \label{fig:cbf}
\end{figure}

The CBF function serves as a projector to convert the \textit{proto-policy} $\pi_\theta$, which is the parameterized actor network from which unsafe actions  might cause congestion, into a \textit{safe policy} $\pi^{CBF}$. Its main objective is to guarantee that the \acrshort{mlu} $\mu$ remains below $1$ (i.e., $\mu(\gls{acbf}) \leq 1$, $\forall \gls{acbf} = \pi^{CBF}(.|s)$). 
As illustrated by Figure \ref{fig:cbf}, this  projection maintains the safe policy $\pi^{CBF}$ as close as possible to the proto policy and keeps load balancing system in safety condition. Safe policy projection is performed at the centralized controller in both learning and testing. Following the safe policy $\pi^{CBF}$, 
each \acrshort{cbf} action is determined according to the following optimization problem: 
\begin{align} \label{eq:safe}
    \begin{split}
       &\gls{acbf}= \underset{\gls{acbf}_t}{argmin}\left\|\gls{acbf}_t - \gls{arl}\right\|_1 \\
    \textit{s.t.} ~~~~~~~~~&  \gls{acbf} \in \mathcal{A} \\
    & \mu(\gls{acbf}) \leq 1 
    \end{split}
\end{align} 

With respect to reward function design, we highlight that the objective function of problem \eqref{eq:P0} is enough to minimize the average tunnel delay in the network. 
However, as mentioned in Section~\ref{sec:system}, when only  minimizing the average delay, the system may induce large delays for a small number of tunnels due to a high link utilization or congestion, i.e. packet drops to further reduce the average delay. Therefore, in order to make the \acrshort{drl} agent also aware of link utilization, we design our reward function to be the weighted sum of the average delay and the \acrshort{mlu} (i.e., $\mu$) as given by Equation \ref{eq:rew}.

\begin{equation} \label{eq:rew}
    r_t(s_t,a_t)=-\sigma \frac{\sum_{k \in K} d_{k,t}}{\left|K \right|} - (1-\sigma )\mu
\end{equation}
where $\sigma \in [0,1]$ emphasizes the importance of the average tunnel delay over a low \acrshort{mlu}. While this reward cannot guarantee itself a hard safety, it guides \acrshort{rl} agent in learning a policy which is both \acrshort{qos} optimal and safe after convergence. To ensure anytime and hard safety, we explain in the following how we incorporate a \acrshort{cbf} function on top of \acrshort{drl} algorithms to constrain the exploration and exploitation so that the \acrshort{mlu} does not exceed 1 (i.e., $\mu\leq 1$).

\textbf{CBF function.} The \acrshort{cbf} function, applied on top of \acrshort{drl} algorithms, brings two main benefits. Firstly, it guarantees that any action, which are stochastically sampled from the parameterized actor network, will not cause a during both exploration and exploitation. Secondly, it considerably contributes to obtaining a better reward-per-learning-step as the \acrshort{mlu} is part of the reward. Indeed, by correcting "bad" proto-actions, which produce a high \acrshort{mlu} and overloaded links, the output safe action  results in a better or at least equal \acrshort{mlu} compared to the initial proto-action. It leads to a fast convergence towards an optimal safe policy. For the projection of proto-actions into safe actions, we propose to use as a \acrshort{cbf} function the local search algorithm detailed in Algorithm \ref{alg:Wapprox_exh}.

In principle, given a proto-action, the \acrshort{cbf}  stochastically attempts to generates $N$ safe actions within a neighborhood of radius $\delta_s$. The generation of actions is based on three different policies where the information with regards to the link utilization of each path $p$ in the tunnel $k$ is exploited:

\begin{algorithm}[tp]
\small
\caption{\acrshort{cbf} based on Local Search algorithm}\label{alg:Wapprox_exh}
Returned action from RL agent $a_{\theta}^{RL}$ \\
Local search radius $\delta_s$ \\
CBF solutions $N$ \\
Local search policy $\pi^{CBF}$\\
Local search max iteration $M$\\
Feasible solution $feas\_sol=\{\}$ \\
\eIf{$\mu(a_{\theta}^{RL}) \leq 1$}{$\gls{acbf}=a_{\theta}^{RL}$}
{
\For{$m$ in $M$}{
\For {$n$ in $N$}{
Randomly sample $\epsilon _n \sim Uniform(0,\delta_s)$\\
Stochastic action generation $a_n^{CBF} \sim \pi^{CBF}$\\
Perform action update: $a_n^{CBF} = a_{\theta}^{RL} \pm \epsilon _n $ \\
\If{$\mu(a_n^{CBF})  \leq 1$}{Append $a_n^{CBF}$ to $feas\_sol$}
}
}
\eIf{$feas\_sol \neq \oslash $}{$\gls{acbf}=\underset{a_i^{CBF} \in feas\_sol}{argmin}\left\|a_i^{CBF} - a_{\theta}^{RL} \right\|_1$}
{$\gls{acbf}= \underset{n}{argmin}\quad \mu\{a_n^{cbf}\}$}
}
\textbf{End}
\end{algorithm}
    
\begin{itemize}
    \item \textbf{\textit{Naive policy}}: This policy randomly picks any tunnel $k \in K$. For each selected tunnel, a random value $\epsilon \sim Uniform(0,\delta_s)$ is added/subtracted to current split ratios given by the proto-action on each path $p \in \gls{pk}$. In particular, the traffic load on the highest utilization path $p$ of the selected tunnel $k$ will be reduced by an amount of $\epsilon.T^k_{t,p}$. This traffic amount will be equally distributed to the remaining paths of the tunnel. In our scenario where there are only 2 paths per tunnel, the reduced traffic on the path with higher utilization will be directly transferred to the path with lower utilization.
    \item \textbf{\textit{DeltaUtil policy}}: This policy selectively focuses on tunnels where the difference between their highest path utilization and lowest path utilization is greater than a certain threshold. In our case, a threshold on the difference of 50 $\%$ is chosen.  Once this criteria is met, a randomly generated value $\epsilon \sim Uniform(0,\delta_s)$ is used to fine-tune split ratios in the proto-action, similarly to the \textbf{Naive} policy.
    \item \textbf{\textit{MaxUtil policy}}: This new policy inherits the principles of the \textbf{\textit{DeltaUtil policy}} policy, but it uses a different criteria for selecting tunnels. 
    Specifically, any tunnel $k$ that has a path load utilization above a threshold of 100 $\%$ (e.g., $ \exists p \in \gls{pk} ~ | ~\mu_p \geq 1$, which is unsafe) will be the target for proto-action modification. Once the set of congested tunnels is determined, each random value $\epsilon_k \sim Uniform(0,\delta_s)$ will be used to adjust the initial split-ratio of each selected tunnel $k$, which is primarily decided by the proto-policy. As a consequence, an amount of traffic $\epsilon_k.T^k_{t,p}$ will be withdrawn from the path $p$ of tunnel $k$ with the highest path utilization, and added to remaining paths.
\end{itemize}

After generating a large number of actions around a proto action, a feasible action (i.e., \acrshort{mlu} is below $100 \%$) is selected 
in such a way that its distance to the original proto action is the smallest. The returned action is quasi-guaranteed to be safe and helps learning safe policies. As local search policies are heuristic, it may be that they cannot correct a proto-action for which a safe action exists. In this case, the \acrshort{cbf} action, which returns the lowest \acrshort{mlu}, will be selected.

Our local search algorithm is important in both training and exploitation phases. In the training phase, it helps to ensure that the network learns safe policies by eliminating the possibility of taking actions that could cause safety issues (i.e., violation of link capacity constraint). In the exploitation phase, it helps to ensure that the network continues to operate safely even under unseen network conditions.

\section{Results and discussions}
\label{sec:results}

We now evaluate the performance of the DRL-CBF solution on the \acrshort{sdwan} scenario presented in Section~\ref{sec:system}.

\subsection{Network environment}

Our simulations are carried out using Python's Gym toolkit \cite{openaigym}, which is a well-known \acrfull{api} for interfacing between our (safe) learning algorithms and our Python-based \acrshort{sdwan} environment. 

\begin{figure}[ht]
    \centering
    \includegraphics[scale=0.4]{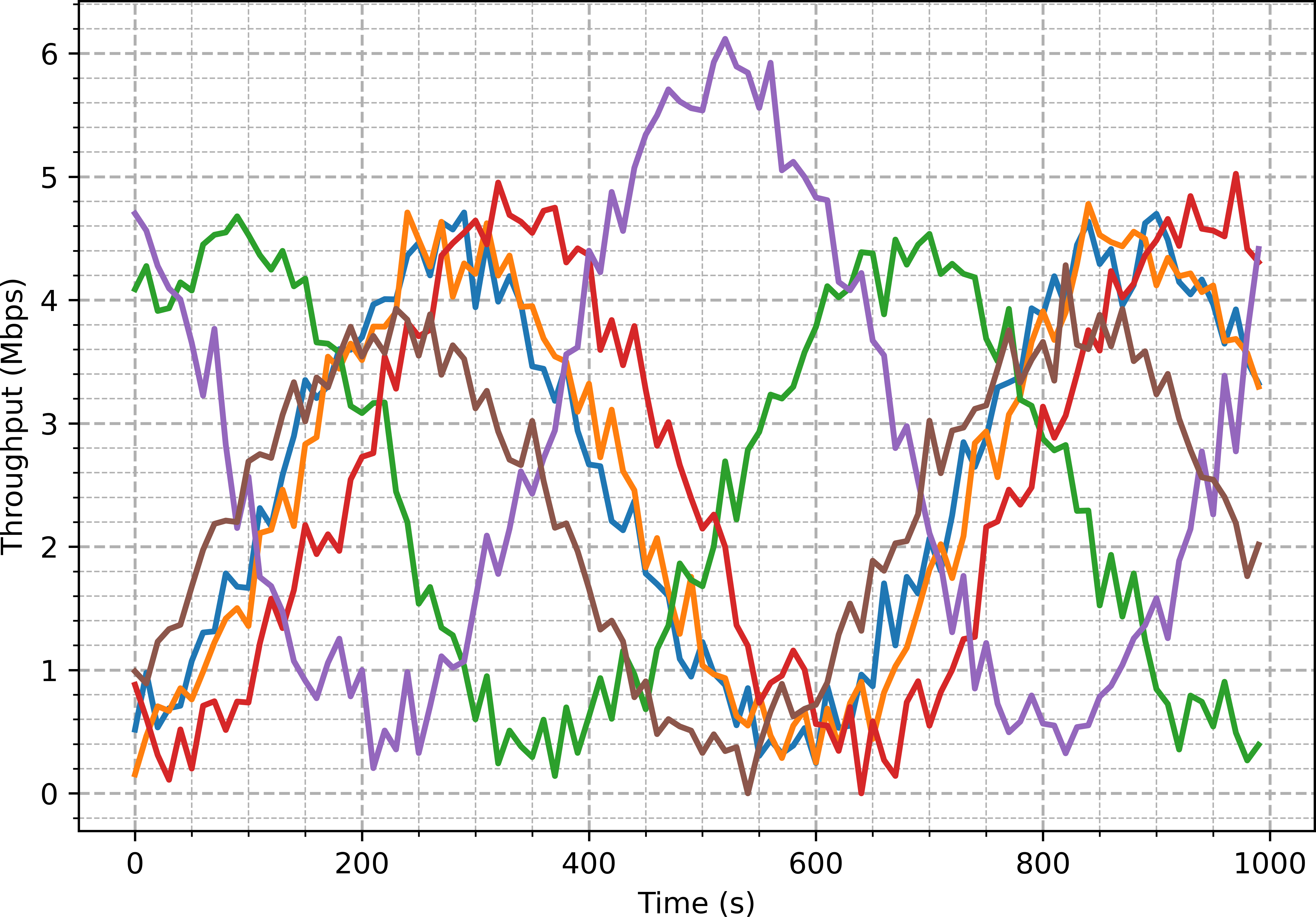}
    \caption{Example of tunnels' traffic over a window of 1000s. 
    }
    \label{fig:traffic}
\end{figure}

In order to demonstrate the stochastic traffic behavior at each \textit{OD flows}, we generate noisy sinusoidal traffic to model diurnal variations, as shown by Figure \ref{fig:traffic}. The phase of each OD flow is shifted to make incoming traffic to each site somewhat variable and likely to cause congestion at  bottleneck links without appropriate load balancing. 

We adopt a simple M/M/1 queuing model to compute the delay on each \textit{link} $e$ and considers a propagation delay $d_{prop}$. The link delay $\gls{de}$ is then derived as follows:
\begin{equation}
    \gls{de} = d_{prop} + \frac{1}{\gls{ce}-l_e}
\end{equation}
In our scenario, each \acrshort{mpls} and Internet link has a fixed capacity of $\gls{ce}=6Mbps$ and  $\gls{ce}=15Mbps$, respectively and  $d_{prop}$ represents the propagation delay. 

The delay on each path $p$ of a tunnel $k$ (i.e. $\gls{dpk}$) is then calculated as the sum of the delay on the edge that is involved in that path as follows:
\begin{equation}
    \gls{dpk} = \sum_{e \in p, p\in \gls{pk}} \gls{de}
\end{equation}

Furthermore, in order to emulate the behavior of \acrshort{tcp} in the considered network, we adopted a min-max fairness rate allocation policy using a standard water-filling algorithm \cite{bertsekasDataNetworks1992}.

\subsection{Algorithms implementation}

Relying on Stable-Baseline3 \cite{stable-baselines3}, a well-known library of \acrshort{rl} algorithms, we have exploited two different types of \acrshort{rl} algorithms: off-policy with \acrshort{ddpg}~\cite{lillicrapContinuousControlDeep2019}) and on-policy with \acrshort{ppo}~\cite{schulmanProximalPolicyOptimization2017}). Given that they do not guarantee  safety, we have applied the \acrshort{cbf} function based on local search algorithms (Algorithm \ref{alg:Wapprox_exh}) on top of them.

\begin{figure}[ht]
    \centering
    \includegraphics[scale=0.4]{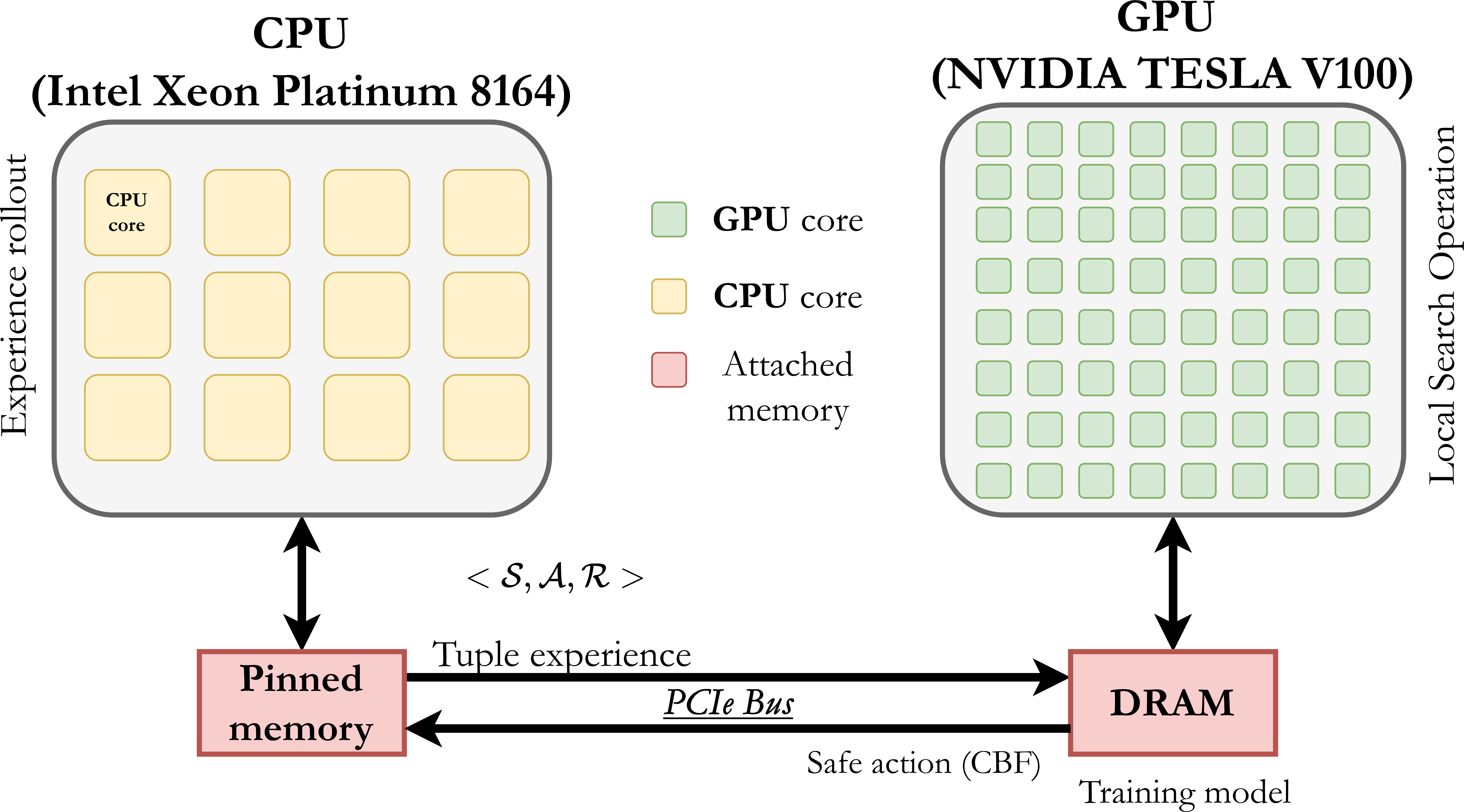}
    \caption{System architecture with 1) network environment running on CPU and 2) safe RL algorithms on GPU.}
    \label{fig:flow}
\end{figure}

We implemented the solution on a server composed by a \acrshort{cpu} Intel\textsuperscript{\textregistered} Xeon\textsuperscript{\textregistered} Platinum 8164 (104 logical cores and 1024 GB of RAM memory) and a \acrshort{gpu} NVIDIA\textsuperscript{\textregistered}  Tesla V100 (5120 \acrshort{cuda} cores and 32 GB of DRAM), as shown in Figure \ref{fig:flow}. As local search algorithms can be massively parallelized, we implemented them with \acrshort{cuda} libraries so that they fully benefit from all \acrshort{gpu} cores available. Therefore, model training and intensive local search algorithms are fully performed at \acrshort{gpu} side. Once a safe policy is found, it is transferred to the \acrshort{cpu}, where our \acrshort{sdwan} environment is located, to perform a rollout step. The resulted tuple experiences are then moved back to \acrshort{gpu} for further model training.

\begin{table}[t]
\centering
\caption{Simulation (hyper)parameters}
\begin{tabular}{c|cc} \label{tab:simul}
                                                                                                    & \multicolumn{2}{c}{\textbf{Value}}                                                                                                                                       \\ \cline{2-3} 
\multirow{-2}{*}{\textbf{\begin{tabular}[c]{@{}c@{}}Simulation \\ (Hyper) parameters\end{tabular}}} & \multicolumn{1}{c|}{\textbf{\begin{tabular}[c]{@{}c@{}}Off-policy Algo\\ (DDPG)\end{tabular}}} & \textbf{\begin{tabular}[c]{@{}c@{}}On-Policy Algo\\ (PPO)\end{tabular}} \\ \hline
\multicolumn{1}{l|}{Gradient clipping maximum}                                                      & \multicolumn{1}{l|}{\cellcolor[HTML]{000000}}                                                  & 0.5                                                                     \\ \hline
Clipping parameter   $\varepsilon$                                                                               & \multicolumn{1}{c|}{\cellcolor[HTML]{000000}}                                                  & 0.2                                                                     \\ \hline
Target KL divergence  $\delta_{KL}$                                                                              & \multicolumn{1}{c|}{\cellcolor[HTML]{000000}}                                                  & 0.03                                                                    \\ \hline
Delayed network update rate                                                                   & \multicolumn{1}{c|}{0.05}                                                                      & \cellcolor[HTML]{000000}{\color[HTML]{000000} }                         \\ \hline
Learning rate                                                                                       & \multicolumn{2}{c}{1e-5}                                                                                                                                                 \\ \hline
Discounted factor  $\gamma$                                                                                 & \multicolumn{2}{c}{0.7}                                                                                                                                                  \\ \hline
Hidden layers                                                                                       & \multicolumn{2}{c}{3}                                                                                                                                                    \\ \hline
Dimension of hidden layer                                                                           & \multicolumn{2}{c}{512}                                                                                                                                                  \\ \hline
Batch size  $|B|$                                                                                        & \multicolumn{2}{c}{256}                                                                                                                                                  \\ \hline
Frequency of model updates (steps)                                                                                        & \multicolumn{2}{c}{256}                                                                                                                                                  \\ \hline
Episode length $L_{eps}$ (steps)                                                                                          & \multicolumn{2}{c}{128}                                                                                                                                                  \\ \hline
Max local search iteration ($M$)                                                                       & \multicolumn{2}{c}{20}                                                                                                                                               \\ \hline
Local search solutions ($N$)                                                                       & \multicolumn{2}{c}{1.000}                                                                                                                                               \\ \hline
Local search radius ($\delta_s$)                                                                    & \multicolumn{2}{c}{0.3}                                                                                                                                                  \\ \hline
Reward parameter $\sigma$                                                                           & \multicolumn{2}{c}{0.8}                                                                                                                                                  \\ \hline
Training steps                                                                          & \multicolumn{2}{c}{1.000.000}                                                                                                                                                  \\ \hline
\end{tabular}
\end{table}

Table~\ref{tab:simul} enumerates the list of parameters that we applied. To stabilize the training process in \acrshort{ddpg}, delayed network updates have been used so that the target actor/critic networks are updated less frequently than the main actor/critic by a factor of $0.05$. The clipped version of \acrshort{ppo}~ \cite{schulmanProximalPolicyOptimization2017} has been used with a clipping parameter  $\varepsilon=0.2$ and a target $KL$ divergence  set at $\delta_{KL}=0.03$. To avoid destructively large weight updates, the gradient is also clipped at 0.5. Besides, a fully connected neural network with 3 layers of 512 neurons is used to approximate policy and value functions.  With regards to local search algorithms, we take the best solution out of  $N*M=20.000$ points generated around each unsafe action returned by the \acrshort{drl} agent 
in a radius of $\delta_s=0.3$. Finally, our models are updated after each rollout of 256 steps, which is equivalent to 2 episodes, and the learning process is studied in $t_s=1.000.000$ training steps. 

\begin{figure}[ht]
        \centering
        \begin{subfigure}[b]{0.24\textwidth}
            \centering
  \includegraphics[width=\textwidth]{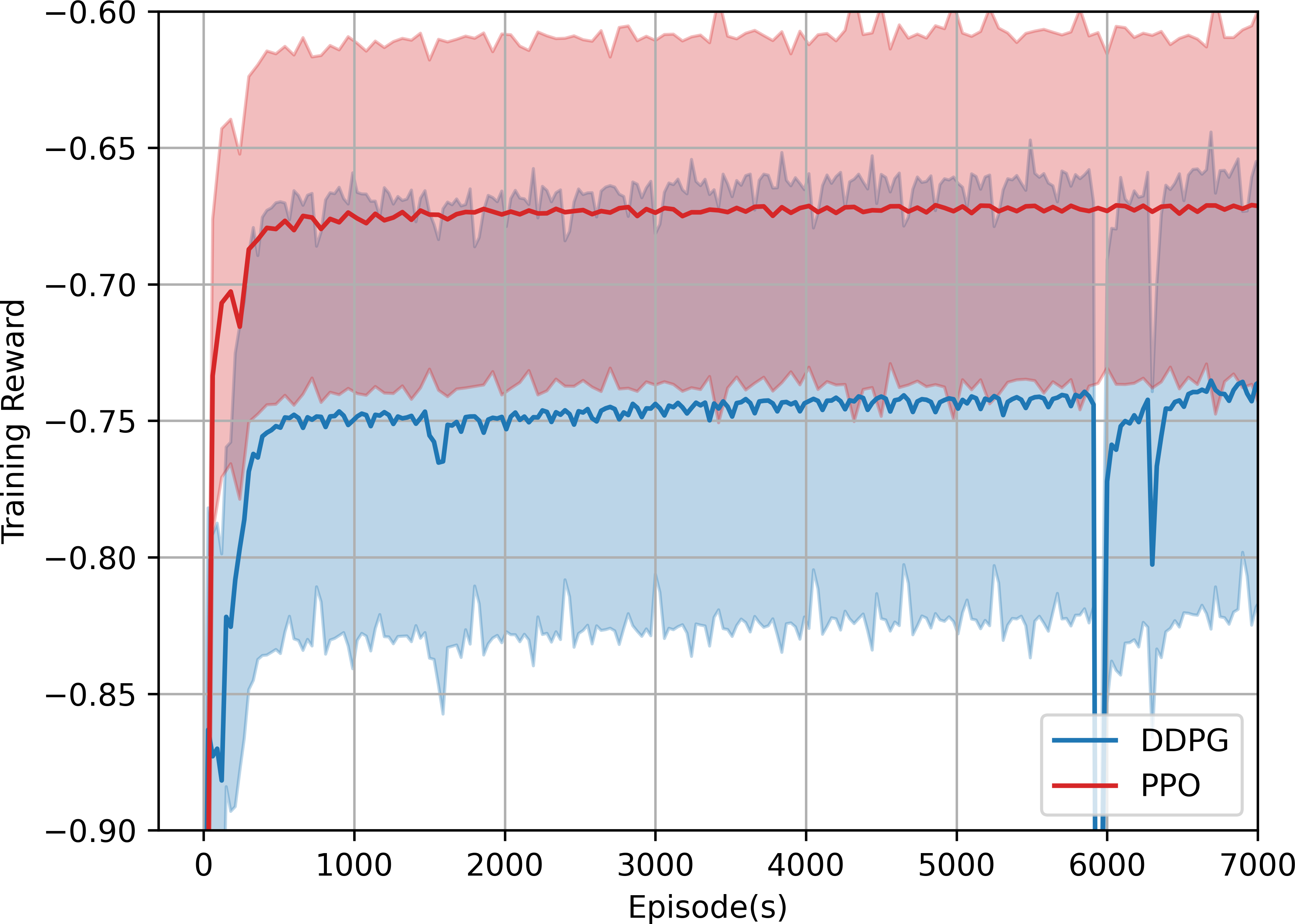}
  \caption{Episodic training rewards}
  \label{fig:rew}
        \end{subfigure}
        \begin{subfigure}[b]{0.23\textwidth}  
            \centering
  \includegraphics[width=\textwidth]{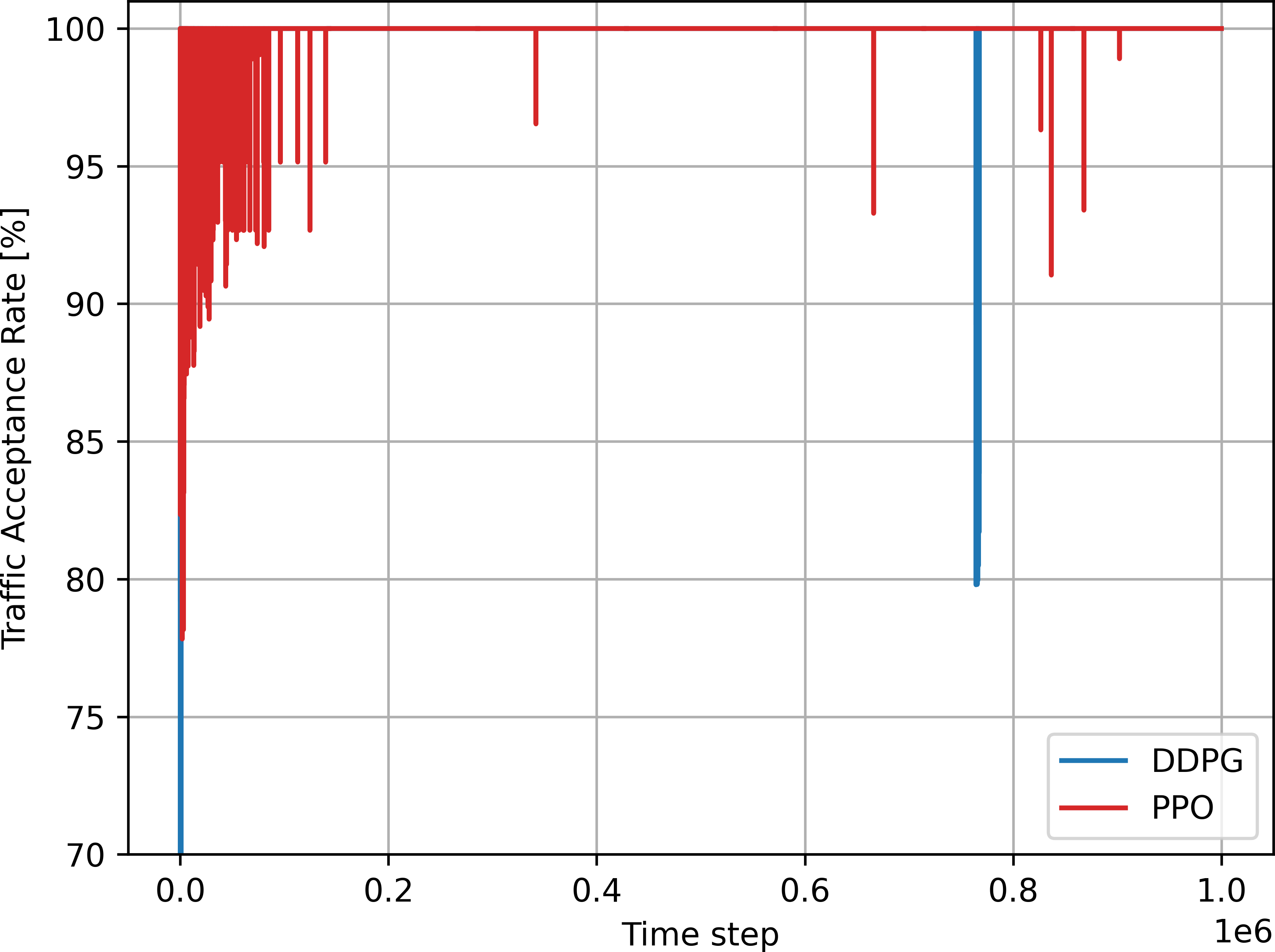}
  \caption{Traffic acceptance rate}
  \label{fig:acr_tr}
        \end{subfigure}
        \caption[ Testing performance results ]
        {\small Episodic training reward and traffic acceptance rate at learning phase.} 
        \label{fig:train_ncbf}
    \end{figure}

To obtain an optimal solution as benchmark, the following constraints \eqref{eq:C2} are added to Problem \ref{eq:P0} so that the link delay is computed according to M/M/1 queuing model. The resulting Non Linear Problem (NLP) is solved by the \acrshort{scip}~\cite{bestuzhevaSCIPOptimizationSuite2021} solver. Note that this constraint plays an important role for our benchmark but, in practice, it is not explicitly known because performance might not follow the M/M/1 model. 

\begin{align*}
     &\gls{de} \geq \frac{1}{c_e - \sum_{k\in K} \sum_{p \in \gls{pk}}\gls{lpk}} \quad \forall e \in E\tag{$\mathcal{C}_2$} \label{eq:C2} \\
\end{align*}

\subsection{Obtained results}

\textbf{On baseline without safety.} Figure \ref{fig:rew} compares the average training reward 
during each episode  for \acrshort{ddpg} and \acrshort{ppo}, respectively. This figure highlights that \acrshort{ppo}  typically achieves better rewards than \acrshort{ddpg}. Besides, the \acrshort{ddpg} agent tends to be trapped into local optimum policies, which can lead to sub-optimal performance.  
In terms of safety, Figure \ref{fig:acr_tr} illustrates the percentage of the total traffic  that is accepted due link capacity constraints.
Both algorithms encourage policy exploration at early stages in training, which accidentally causes link congestion and traffic rejection. A severe unsafe attempt of policy exploration for DDPG can be observed at episode 6000. However, even if it helps learning better policies, 
unsafe exploratory actions should be avoided in production systems, especially for on-policy algorithms where the model is updated and applied after a smaller number of episodes.

\begin{figure*}[htp]
        \centering
        \begin{subfigure}[b]{0.325\textwidth}
            \centering
    \includegraphics[width=\textwidth]{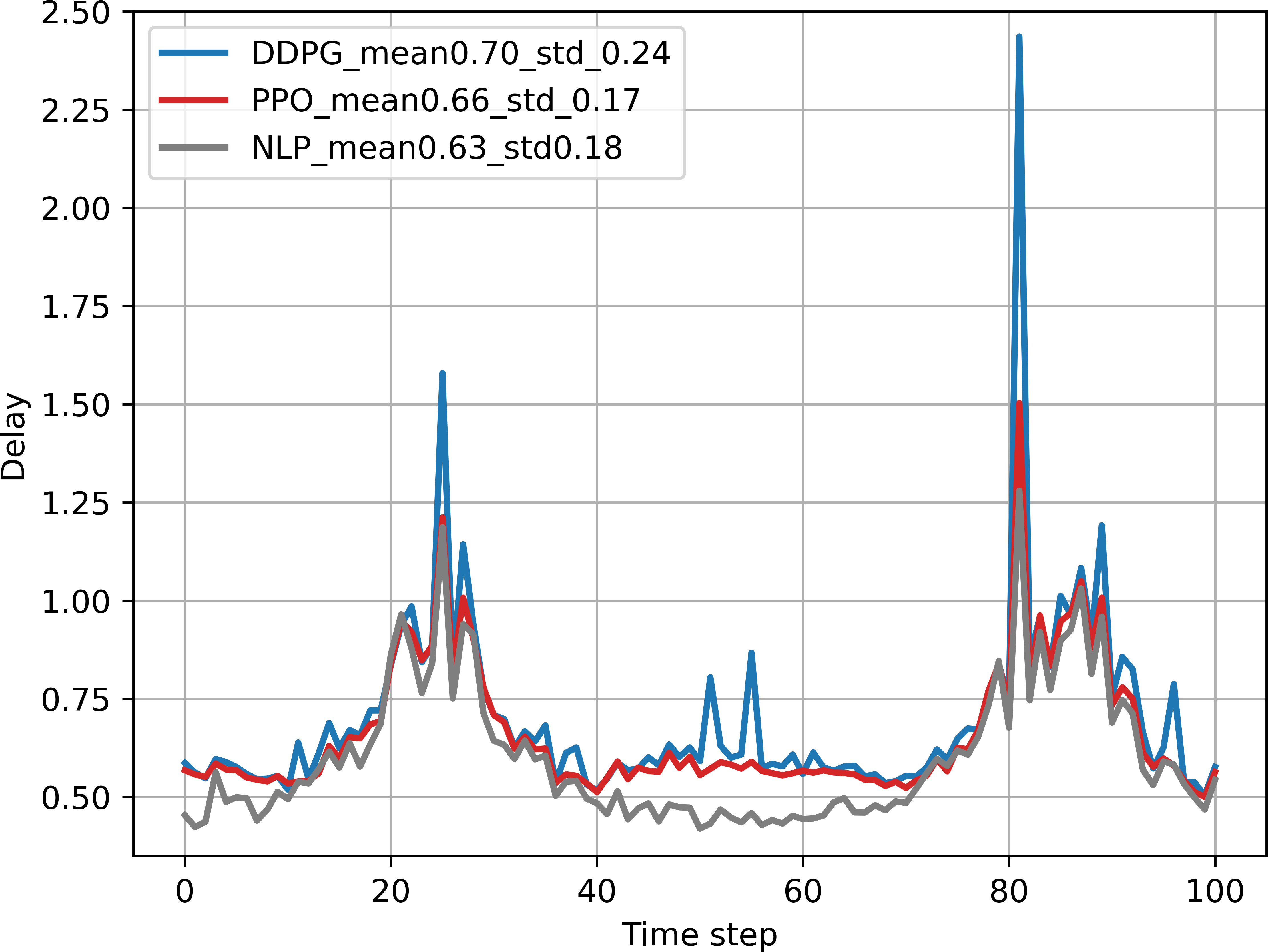}
    \caption{Delay results of DDPG, PPO and NLP}
    \label{fig:del}
        \end{subfigure}
        \begin{subfigure}[b]{0.325\textwidth} 
            \centering
    \includegraphics[width=\textwidth]{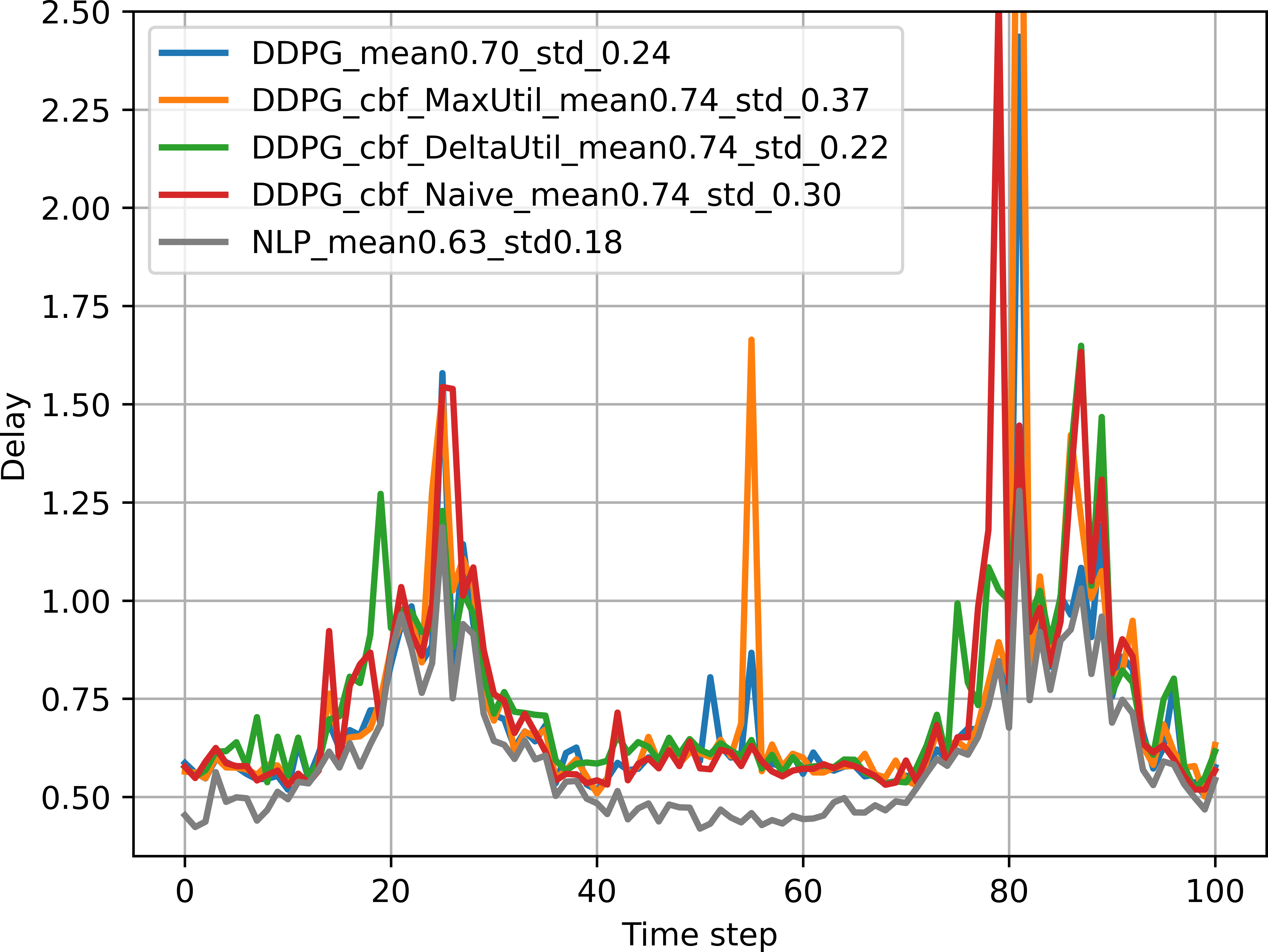}
    \caption{Delay results of  DDPG-CBF and NLP}
    \label{fig:delddpg}
        \end{subfigure}  
        \begin{subfigure}[b]{0.325\textwidth} 
           \centering
    \includegraphics[width=\textwidth]{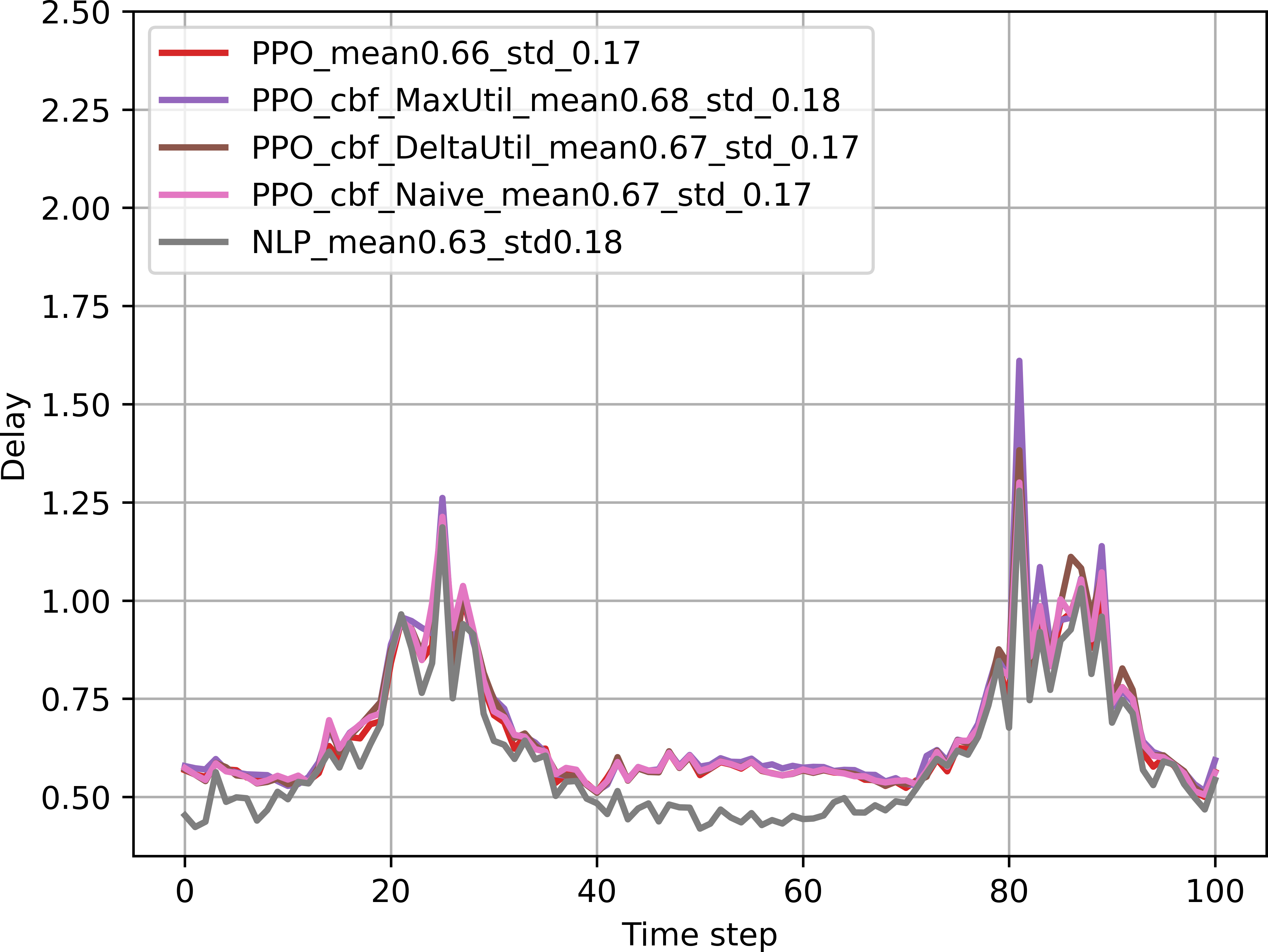}
    \caption{Delay results of PPO-CBF and NLP}
    \label{fig:delppo}
        \end{subfigure}
        % \hfill  
        \vskip\baselineskip
        % \hfill 
        \begin{subfigure}[b]{0.325\textwidth} 
            \centering
    \includegraphics[width=\textwidth]{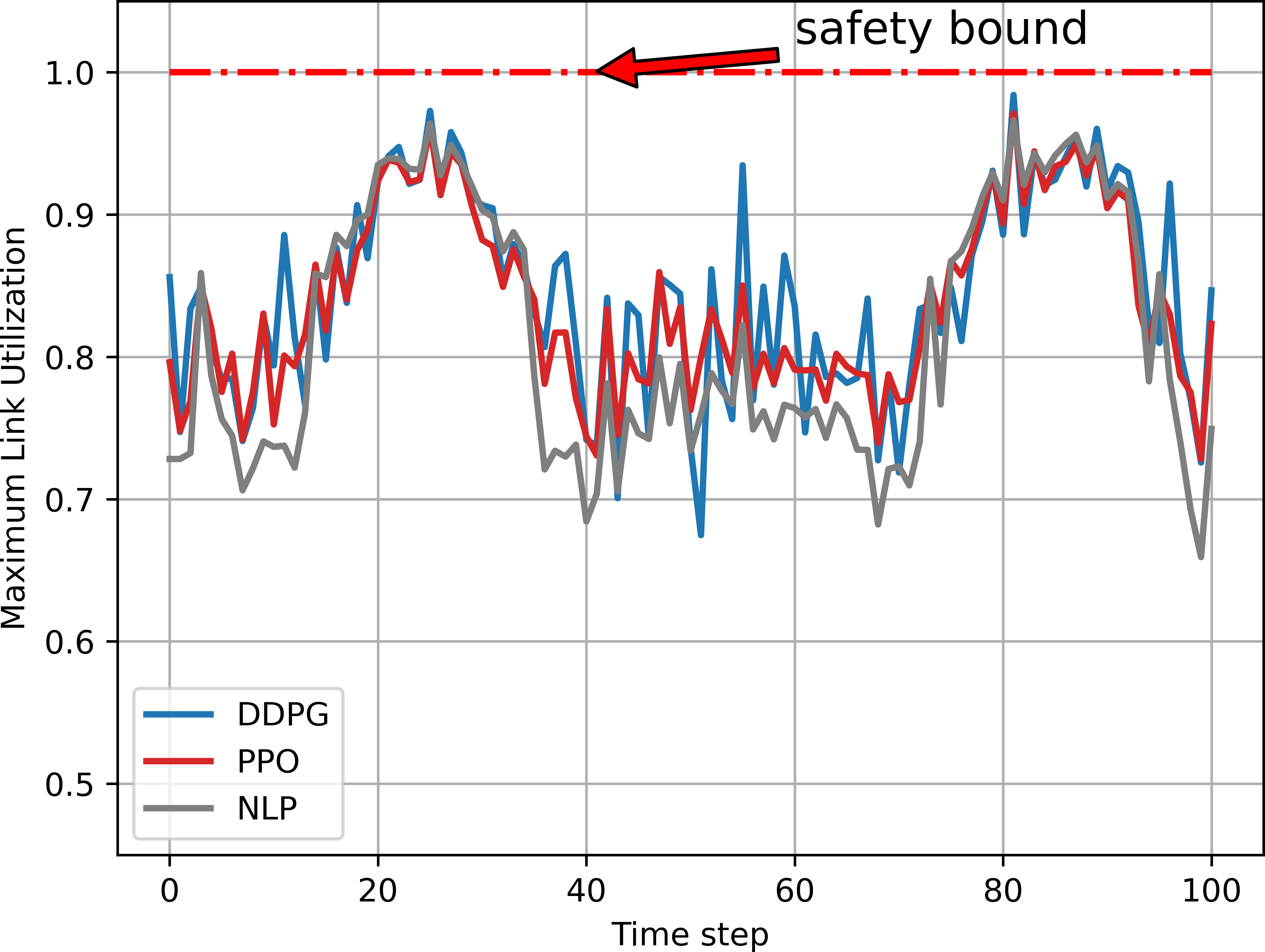}
    \caption{MLU results of DDPG, PPO and NLP }
    \label{fig:mlu}
        \end{subfigure}  
        \begin{subfigure}[b]{0.325\textwidth} 
           \centering
    \includegraphics[width=\textwidth]{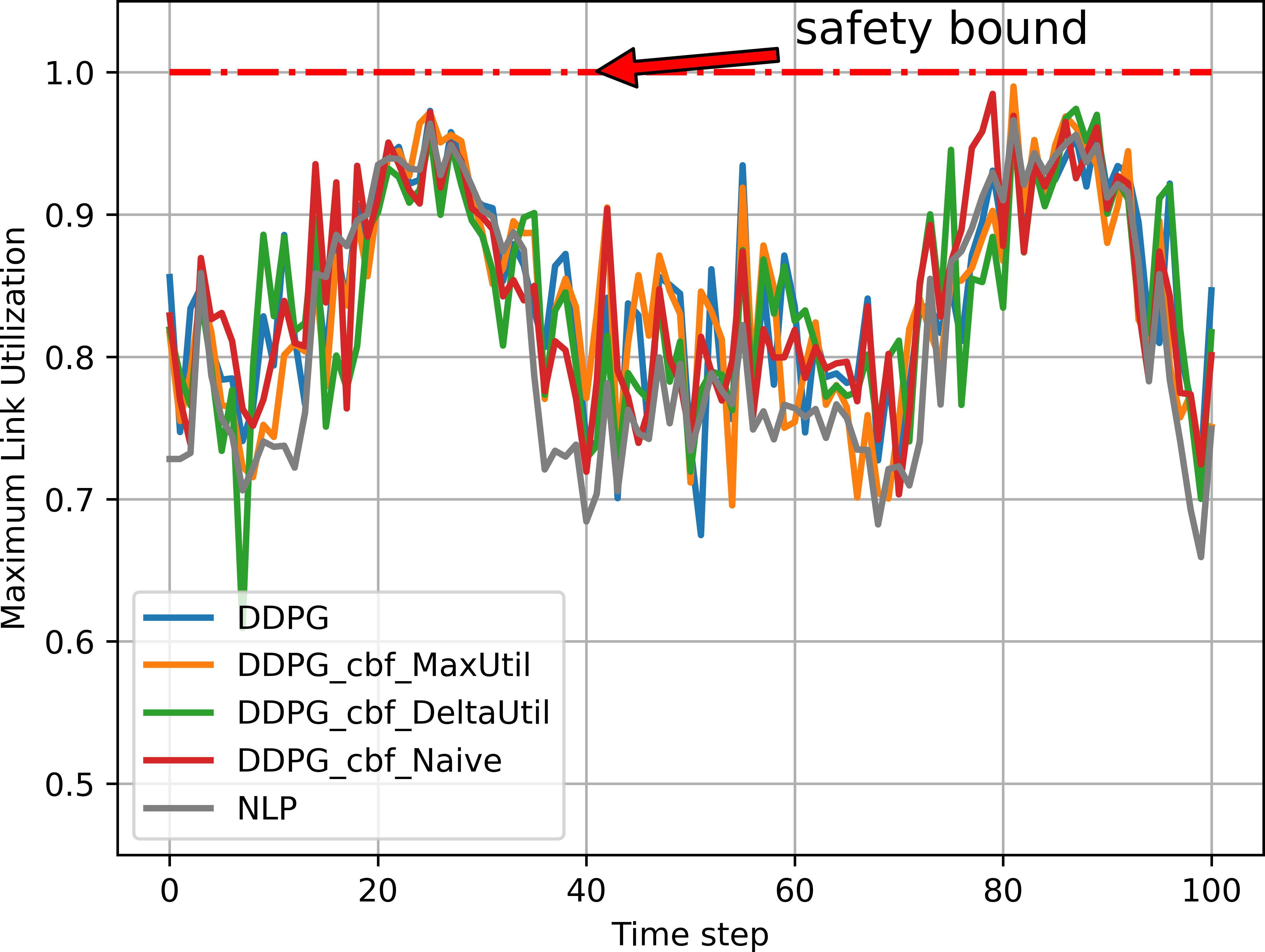}
    \caption{MLU results of DDPG-CBF and NLP}
    \label{fig:mluddpg}
        \end{subfigure}       
        \begin{subfigure}[b]{0.325\textwidth} 
            \centering
  \includegraphics[width=\textwidth]{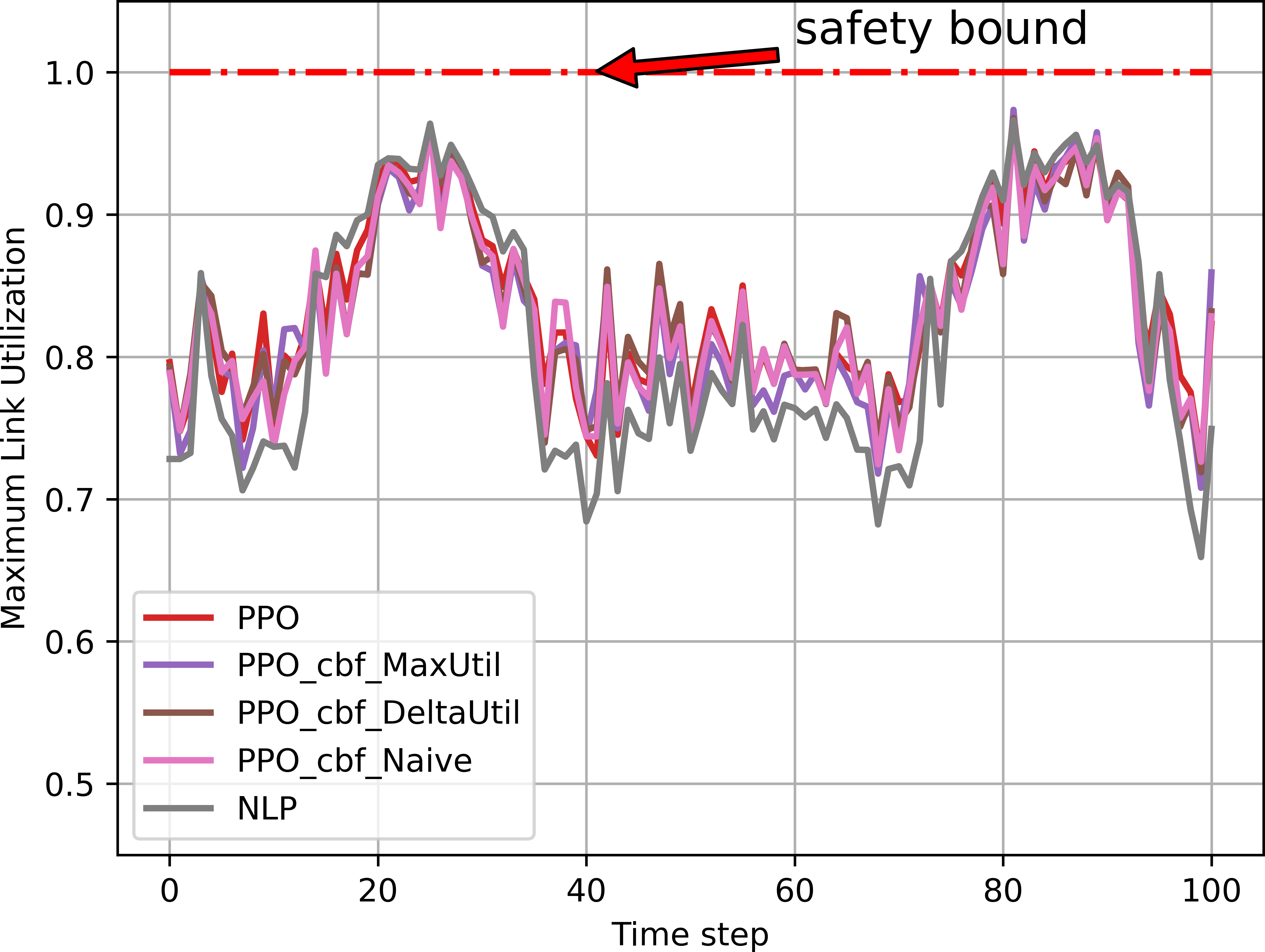}
  \caption{MLU results of PPO-CBF and NLP}
  \label{fig:mluppo}
        \end{subfigure}        
        \caption[ Local traffic and corresponding shaping rate, loss of each solution ]
        {\small Testing performance on average tunnel delay and MLU.} 
        \label{fig:train_test}
\end{figure*}

In order to compare performance during testing, our learning models using \acrshort{ddpg} and \acrshort{ppo} algorithms are selected after the last episode training. At this point, training models are well aware of delay and \acrshort{mlu} minimization in the objective function. Therefore, traffic rejection due to link capacity violation is unlikely to happen. To define a testing scenario, we generated 100 traffic samples for each OD tunnel following the traffic pattern presented in Figure \ref{fig:traffic}. 
As illustrated in Figures \ref{fig:del} and \ref{fig:mlu}, delay and \acrshort{mlu} of learning-without-safety models are compared to the optimal \acrshort{nlp} solution obtained with \acrshort{scip}. They reveal that near-optimal delays are obtained using conventional \acrshort{ddpg} and \acrshort{ppo} learning algorithms. Besides, the \acrshort{mlu} during testing is safely kept below $1$, resulting in no traffic rejection. Furthermore, network delay is better for \acrshort{ppo} compared to \acrshort{ddpg}  since the local optimum trap is not observed during training.

\begin{figure*}[htp]
        \centering
        \begin{subfigure}[b]{0.251\textwidth}
            \centering
  \includegraphics[width=\textwidth]{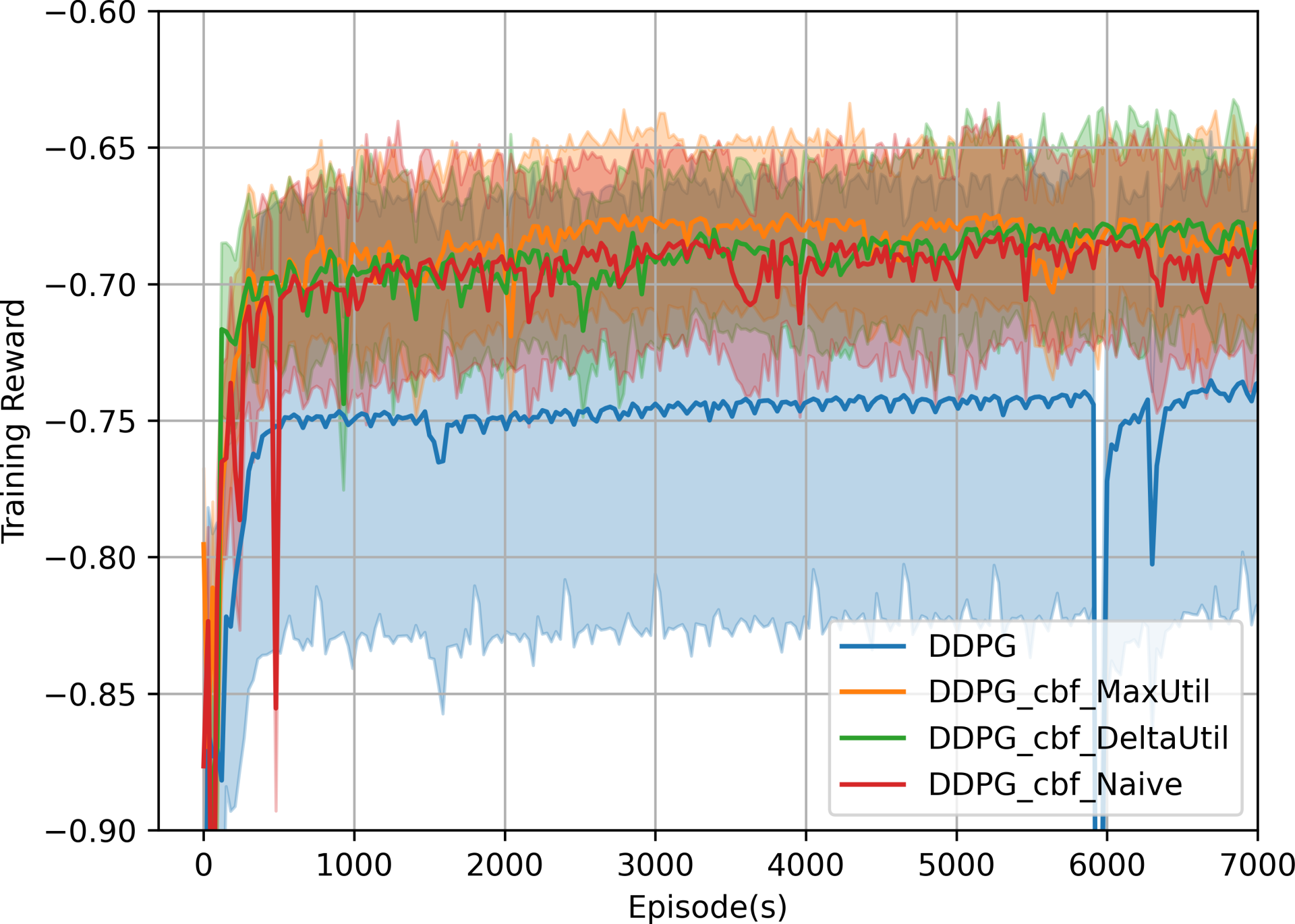}
  \caption{Reward of DDPG-CBF}
  \label{fig:rew_ddpg}
        \end{subfigure}
        % \hfill
        \begin{subfigure}[b]{0.251\textwidth}  
            \centering
  \includegraphics[width=\textwidth]{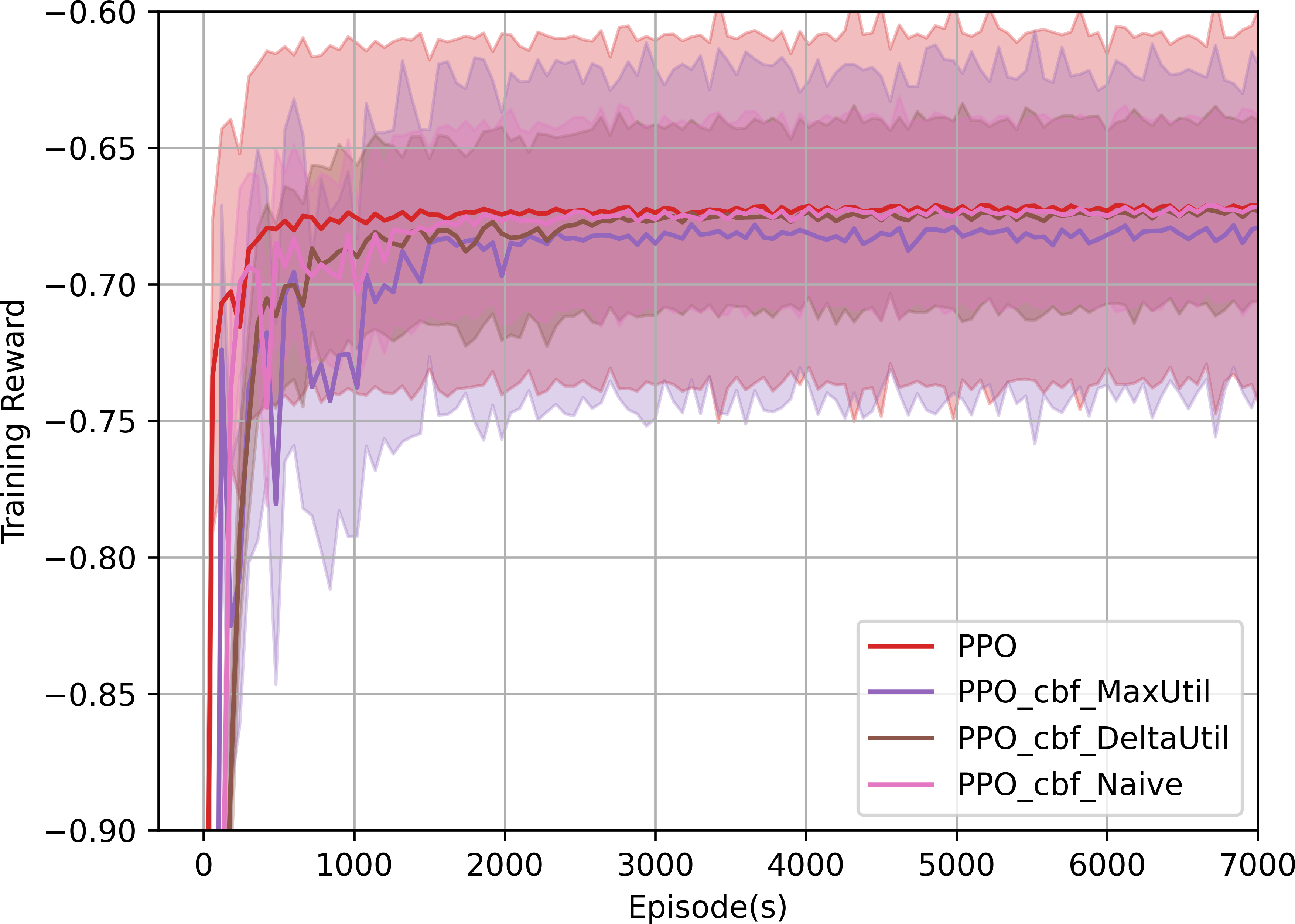}
  \caption{Reward of PPO-CBF}
  \label{fig:rew_ppo}
        \end{subfigure}
                \begin{subfigure}[b]{0.24\textwidth} 
           \centering
    \includegraphics[width=\textwidth]{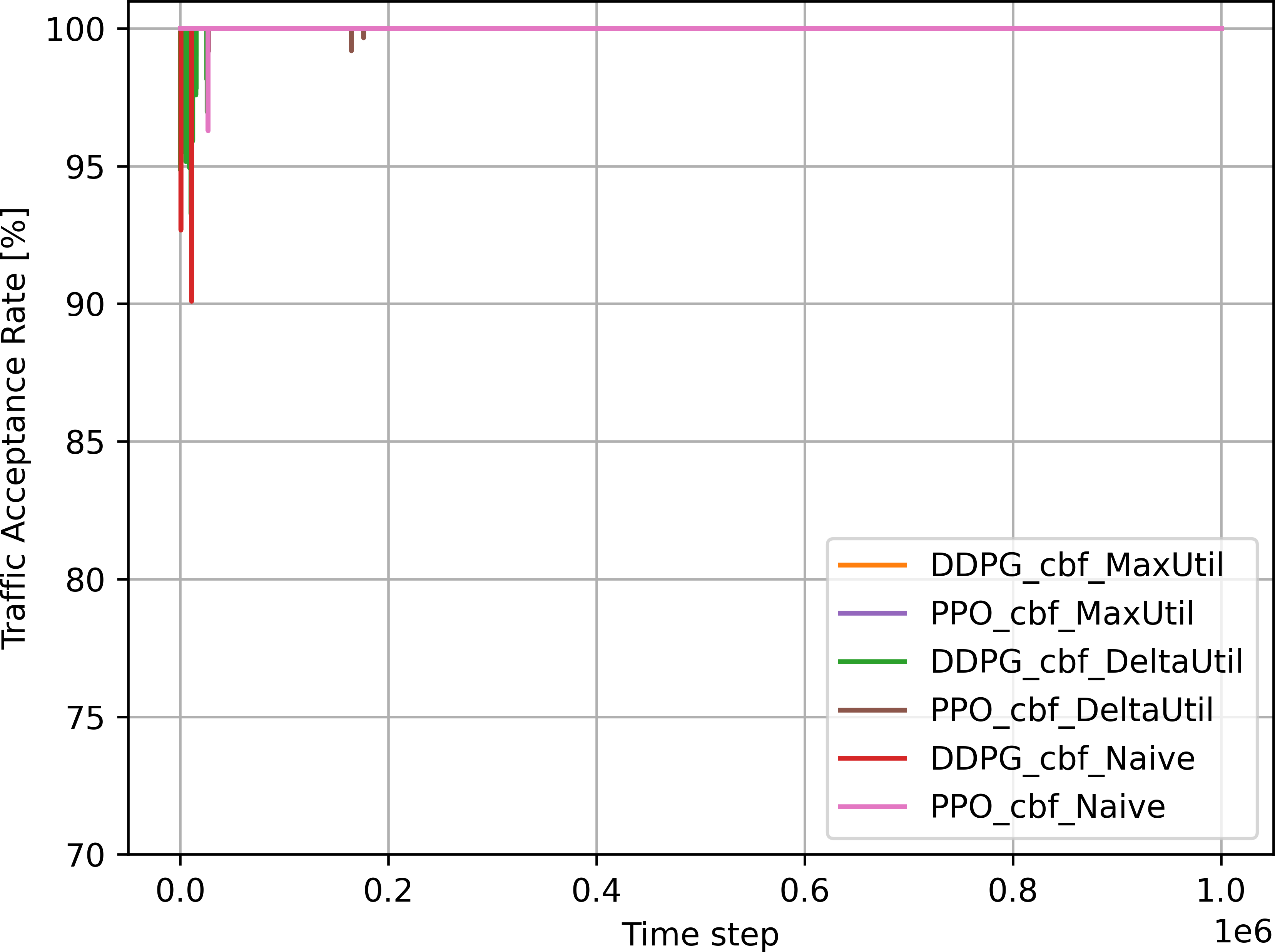}
    \caption{Traffic acceptance rate}
    \label{fig:acr_tr_cbf}
        \end{subfigure}
        % \hfill     
        \begin{subfigure}[b]{0.24\textwidth} 
            \centering
  \includegraphics[width=\textwidth]{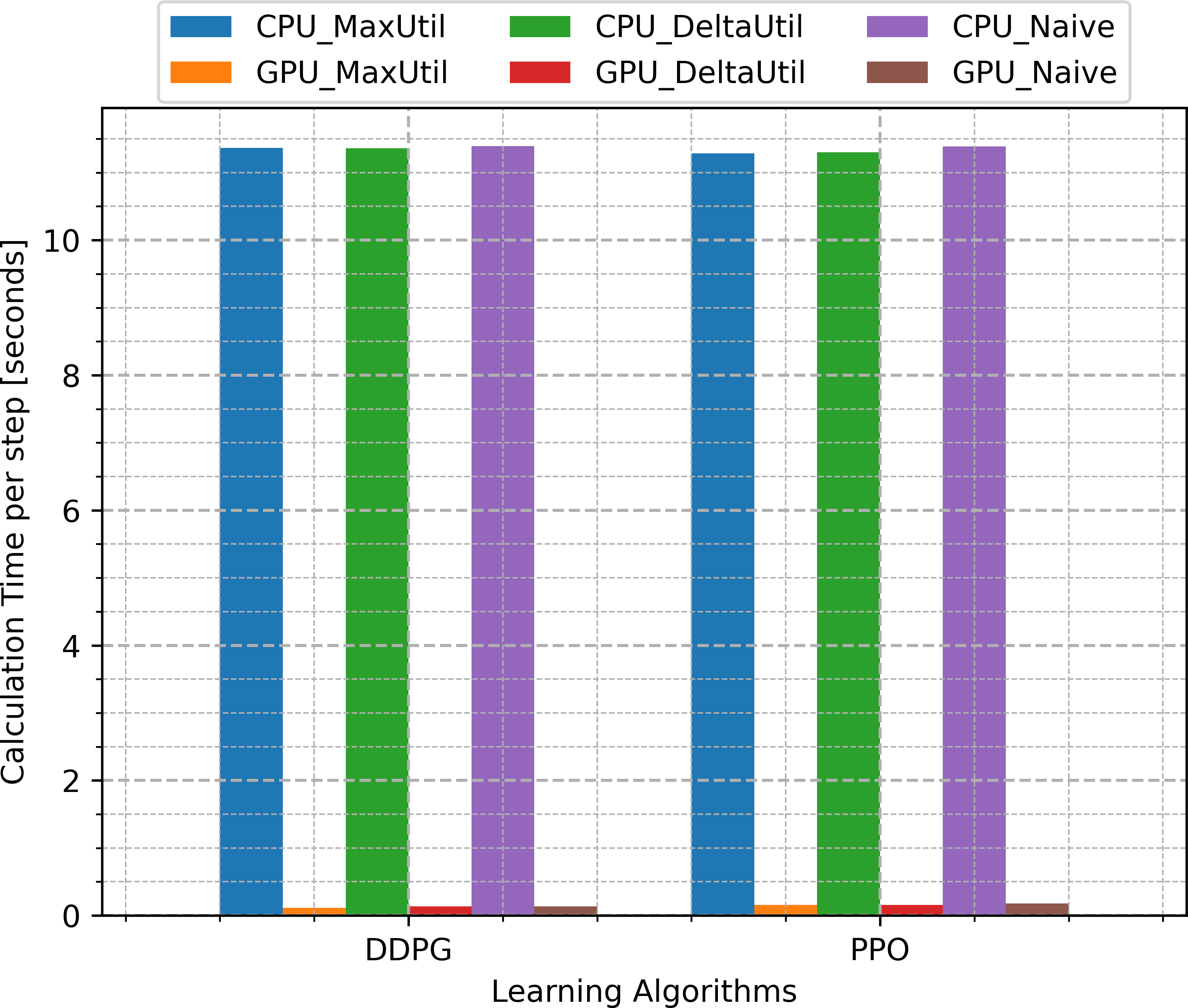}
  \caption{Calculation time}
  \label{fig:calcul_time}
        \end{subfigure}
        \caption[ Training performance results ]
        {\small Training performance: (a,b) average episodic training reward, (c) traffic acceptance rate and (d) average computation time per step.} 
        \label{fig:training_cbf}
\end{figure*}

\textbf{With safety.} When the safety \acrshort{cbf} layers are applied on top of current off/on policy learning algorithms, testing performance results %and traffic acceptance rate during learning 
are depicted in %\jeremie{should be bcef?} \lam{fixed} 
Figures \ref{fig:delddpg}, \ref{fig:mluddpg}, \ref{fig:delppo} and \ref{fig:mluppo}, respectively. With respect to safety, off-policy learning using \acrshort{ddpg}, Figures \ref{fig:delddpg} and \ref{fig:mluddpg} demonstrate that the \acrshort{ddpg}-\acrshort{cbf} agent does not handle delay well, especially when the total traffic demand is high. In particular, many high delay peaks are observed during testing phase regardless \acrshort{cbf} policies, although safety is always respected (\acrshort{mlu} belows $1$). 
\acrshort{ppo}-\acrshort{cbf}  significantly improves network delay and better controls the \acrshort{mlu} in testing phase as illustrated by Figures \ref{fig:delppo} and \ref{fig:mluppo}. Furthermore, no capacity violations happen during learning as indicated by Figure \ref{fig:acr_tr_cbf}. These results suggest that, with safety, on-policy learning outperforms off-policy learning as near-optimal performance is obtained and hard safety requirements are met. 

In addition, Figure \ref{fig:acr_tr_cbf} also shows the training performance for the three \acrshort{cbf} functions we implemented. As we can see, traffic acceptance during learning is significantly improved compared to the case without \acrshort{cbf}, as illustrated in Figure \ref{fig:acr_tr}. Moreover, the \textbf{MaxUtil} policy outperforms \textbf{Naive} and \textbf{DeltaUtil}  policies in achieving no traffic rejection during learning. Indeed, this policy directly targets tunnels that make the network congested. With respect to testing performance, Figure \ref{fig:delddpg}, \ref{fig:mluddpg}, \ref{fig:delppo} and \ref{fig:mluppo} show that all three \acrshort{cbf} functions nearly get the same performance and that \acrshort{ppo}-\acrshort{cbf} performs much better than \acrshort{ddpg}-\acrshort{cbf}. Indeed, no delay spikes and smoother performance are observed for \acrshort{ppo}-\acrshort{cbf}.

To shed the light on the convergence of training rewards with and without \acrshort{cbf} functions, Figure \ref{fig:rew_ddpg} and \ref{fig:rew_ppo} compare the episodic training rewards of \acrshort{ddpg}-\acrshort{cbf} and \acrshort{ppo}-\acrshort{cbf}, respectively. These figures reveal that on-policy learning is faster and stable compared to off-policy learning. %\jeremie{broken:} 
Although \acrshort{ddpg}-\acrshort{cbf} achieves a better reward than its non-safe version, its training curve is slightly unstable compared to the training curves of \acrshort{ppo}-\acrshort{cbf}. As a result, combining safe learning with this off-policy approach is more challenging.

Figure \ref{fig:calcul_time} demonstrates the benefits of our hardware architecture illustrated on Figure \ref{fig:flow} for training acceleration. %\jeremie{with PPO?}. 
We compare the average, real execution time required to successfully perform one iteration of Algorithm \ref{alg:rlAlgo}, which includes local search algorithm and model update, with and without (i.e. \acrshort{cpu}) using the \acrshort{gpu}. It can be observed that the \acrshort{gpu} implementation significantly speeds up calculation and training time. In particular, a feasible (safe) action is found in less than $0.1(s)$ when using \acrshort{gpu}, compared to more than $11(s)$ using \acrshort{cpu}. This significant time saving favors the usage of our methods in practice, since  the model can also be updated every 256 steps in roughly 25.6 seconds, rather than 2816 seconds using only \acrshort{cpu}. As a result, our model can be fully trained in one day using 1.000.000 training steps instead of 134 days when using the \acrshort{cpu}. This time acceleration has a significant impact on the possibility to deploy our solution in practice. Indeed, since network measurements (e.g., delay, traffic, loss, etc.) can be collected every $1$s, on-policy leaning models can be updated every $256$s in $25$s,
which is reasonable.

\section{Conclusion}
\label{sec:conclusion}
We presented a novel approach combining the \acrfull{drl} and a \acrfull{cbf} to guarantee safe exploration and exploitation in the context of \acrfull{sdwan}. Tackling a typical load balancing problem where latency needs to be  optimized while meeting safety requirements in terms of capacity constraints, our \acrshort{drl}-\acrshort{cbf} approach is able to achieve near-optimal performance with safe exploration and exploitation. Furthermore, we show that on-policy optimization based on \acrshort{ppo} achieves better performance than off-policy learning with \acrshort{ddpg}. We implemented all the algorithms on GPU to accelerate training by approximately 110x times and achieve model updates for on-policy methods within a few seconds, making the full solution practical.

Future works along these lines include the integration with a network simulator and a testbed for a more realistic performance evaluation. We also plan to address more challenging environments where, for instance, QoE needs to be optimized and other constraints need to be handled.

\section*{Acknowledgment}
This work was partially supported by the French Nation Research Agency (ANR) SAFE project  under grant ANR-21-CE25-0005.

\bibliographystyle{IEEEtran}
\bibliography{biblio.bib}

\vspace{12pt}

\end{document}